          [2001/04/25 1.1 (PWD)]
\documentclass{aa}

\usepackage{graphicx,epsfig}
\usepackage{natbib}
\usepackage{epstopdf}
\bibpunct[, ]{(}{)}{;}{a}{}{,}

\newcommand{\beq}{\begin{equation}}
\newcommand{\eeq}{\end{equation}}
\newcommand{\beqa}{\begin{eqnarray}}
\newcommand{\eeqa}{\end{eqnarray}}

\begin{document}

\titlerunning{Non-linear Alfv\'en Waves in Magnetic Flux Tubes}
\title{Torsional Alfv\'en Waves in Solar Magnetic Flux Tubes of Axial Symmetry
}

\author{
K. Murawski \inst{1}
\and
A. Solov'ev \inst{2}
\and 
Z.E. Musielak \inst{3,4}
\and
A.K. Srivastava \inst{5} 
\and
J. Kra\'skiewicz \inst{1}  
}

\offprints{K. Murawski}

\institute{Group of Astrophysics, 
      University of Maria Curie-Sk{\l}odowska, 
      ul. Radziszewskiego 10, 
      20-031 Lublin, Poland \\
      \email{kmur@kft.umcs.lublin.pl}
\and
      Central (Pulkovo) Astronomical Observatory, Russian 
			Academy of Sciences, St. Petersburg, Russia\\
      \email{solov@gao.spb.ru}
\and
      Department of Physics, University of Texas at Arlington,
      Arlington, TX 76019, USA \\
      \email{zmusielak@uta.edu}
\and
      Kiepenheuer-Institut f\"ur Sonnenphysik, Sch\"oneckstr. 6,
      79104 Freiburg, Germany \\
			\email{musielak@kis.uni-freiburg.de}
\and
      Department of Physics, Indian Institute of Technology (Banaras 
			Hindu University), Varanasi-221005, India \\
			\email{asrivastava.app@iitbhu.ac.in}
}

\date{Received <date> / Accepted <date>}

\abstract
% context heading (optional)
{}
% aims heading (mandatory)
{Propagation and energy transfer of torsional Alfv\'en waves in 
solar magnetic flux tubes of axial symmetry is studied.
}
% methods heading (mandatory)
{
An analytical model of a solar magnetic flux tube of axial symmetry 
is developed 
by specifying a magnetic flux and deriving general analytical formulae 
for the equilibrium mass density and a gas pressure. The main advantage 
of this model is that it can be easily adopted to any axisymmetric 
magnetic structure. The model is used to simulate numerically the 
propagation of nonlinear Alfv\'en waves in such 2D flux tubes of axial symmetry embedded 
in the solar atmosphere.  The waves are excited by a localized pulse in 
the azimuthal component of velocity and launched at the top of the solar 
photosphere, and they propagate through the solar chromosphere, transition 
region, and into the solar corona.}
% results heading (mandatory)
{
The results of our numerical simulations reveal a complex scenario of 
twisted magnetic field lines and flows associated with torsional Alfv\'en 
waves as well as energy transfer to the magnetoacoustic waves that are triggered by the 
Alfv\'en waves and are akin to the vertical jet flows.  Alfv\'en waves experience about 
$5\%$ amplitude reflection at the transition region. 
Magnetic (velocity) field perturbations experience attenuation (growth) with height 
is agreement with analytical findings. Kinetic energy of magnetoacoustic waves 
consists of $25\%$ of the total energy of Alfv\'en waves. 
The energy transfer may lead to localized mass transport in the form of vertical jets, 
as well as to localized heating as slow magnetoacoustic waves are prone to dissipation  
in the inner corona.}
% conclusions heading (optional)
{}

\keywords{Sun: atmosphere -- (magnetohydrodynamics) MHD -- waves}

\maketitle

%%%%%%%%%%%%%%%%%%%%%%%%%%%%%%%%%%%%%%%%%%%%%%%%%%%%%%%%%%%%%%%%%%%%%%%%%%%%%%%%%%%%%%%%%%%

\section{Introduction}

The role of magnetohydrodynamic (MHD) waves in transporting energy and heating 
various layers of the solar atmosphere, and acceleration of the solar wind, has 
been studied by many authors (e.g., Hollweg 1981; Hollweg et al. 1982; Priest 
1982; Hollweg 1992; Musielak 1998; Ulmschneider \& Musielak 1998, 2003; Dwivedi \& 
Srivastava 2006, 2010; Zaqarashvili \& Murawski 2007; Gruszecki et al. 2008; 
Murawski \& Musielak 2010; Ofman 2009; Chmielewski et al. 2013).  Among the 
three basic MHD modes, Alfv\'en waves are of a particular interest as they can 
carry their energy along magnetic field lines to higher atmospheric altitudes 
on fast temporal scales (e.g., Hollweg 1978, 1981, 1992; Musielak \& Moore 1995; 
Cargill et al. 1997; Vasheghani Farahani et al. 2011, 2012). 

The interest in Alfv\'en waves has recently significantly increased because of 
their 
%direct 
observational evidence in the solar atmosphere. Several authors 
claim that they have already observed outwardly-propagating Alfv\'en waves in 
the quiescent solar atmosphere in the localized solar magnetic flux tubes
(e.g., Jess et al. 2009; Fujimura \& Tsuneta 2009; 
%McIntosh et al. 2011; 
Okamoto \& De Pontieu 2011; 
De Pontieu et al. 2012; Sekse et al. 2013), 
while Van Doorsselaere et al. (2008) questioned the interpretation of Alfv\'en wave 
observations in the solar atmosphere. 
%; Tien et al. 2012). 
Specifically, 
Jess et al. (2009) interpreted the H$\alpha$ data in terms of torsional Alfv\'en 
waves in the solar chromosphere, with periods ranging from $2\,$ min to near $12\,$ 
min, and with the maximum power near $6-7\,$ min.  Torsional and swirl-like motions 
were also reported by respectively Bonet et al. (2008) and Wede\-mey\-er-B\"ohm \& 
Rouppe van der Voort (2009).  According to Bonet et al. (2008), there are vortex 
motions of G band bright points around downflow zones in the photosphere, and 
lifetimes of these motions are of the order of $5$ min.  Wede\-mey\-er-B\"ohm \& 
Rouppe van der Voort (2009) observed disorganized relative motions of photospheric 
bright points and concluded that they induce swirl-like motions in the solar 
chromosphere. 

Different theoretical aspects of the generation, propagation and dissipation of 
torsional Alfv\'en waves in the solar atmosphere were investigated by Hollweg 
(1978, 1981), Heinemann \& Olbert (1980), and Ferriz-Mas et al. (1989), who 
solved the Alfv\'en wave equations using different sets of wave variables.  
Musielak et al. (2007) demonstrated that the propagation of these waves along 
isothermal and thin magnetic flux tubes is cutoff-free. However, Routh et al. 
(2007, 2010) showed that gradients of physical parameters, such as temperature, 
are responsible for the origin of cutoff frequencies, which restrict the wave 
propagation to certain frequency intervals.  In most of the above work only 
linear (small-amplitude) Alfv\'en waves were considered. 

Numerical studies of linear and nonlinear torsional Alfv\'en waves in solar 
magnetic flux tubes were performed by Hollweg et al. (1982), who considered 
a one-dimensional (1D) model.  The original work of Holweg et al. was extended 
to higher dimensions by Kudoh \& Shibata (1999) and Saito et al. (2001), who 
investigated the effects caused by nonlinearities.  Fedun et al. (2011) showed 
by means of numerical simulations that chromospheric magnetic fluxtubes can act 
as a frequency filter for torsional Alfv\'en waves. Wede\-mey\-er-B\"ohm et al. 
(2012) used the observational results originally reported by Wede\-mey\-er-B\"ohm 
\& Rouppe van der Voort (2009) to demonstrate that the swirl-like motions observed 
in the lower parts of the solar atmosphere produce magnetic tornado-like motions 
in the solar transition region and corona.  However, more recent studies performed 
by Shelyag et al. (2013) seem to imply that the tornado-like motions actually do 
not exist once time dependence of the local velocity field is taken into account. 
Instead, the tornado-like motions were identified as torsional Alfv\'en waves 
propagating along solar magnetic flux tubes. 
Recently, Wedemeyer-B\"ohm \& Rouppe 
van der Voort (2009) 
%discovered 
reported on 
small, rotating swirls observed in the solar chromosphere, 
and interpreted them 
%their observations 
as plasma spiraling upwards in a funnel-like magnetic structure. 
%Recently, 
Chmielewski et al. (2014) and 
Murawski et al. (2014) 
modeled 
%the fast magnetic twister 
such short-lived swirls 
and associated plasma and torsional motions 
in coronal magnetic 
%arcades. 
structures. 

As a result of the expanding, curved and strong magnetic field lines of solar
magnetic flux tubes embedded in the solar atmosphere, formulating a model of 
such flux tubes is always a formidable task.  Nevertheless, many efforts were 
undertaken in the past to develop realistic flux tube models.  Among many others, 
let us mention Low (1980) and Gent et al. (2013, and references therein), who 
constructed magnetic flux tube models in the solar atmosphere. In particular, 
Low (1980) formulated an analytical sunspot model, while Gent et al. (2013) 
solved analytically the magnetohydrostatic equilibrium problem. In this paper, 
we develop a model of a solar magnetic flux tube of axial symmetry. 
The model requires 
specifying a magnetic flux and gives analytical formulae for the equilibrium 
mass density and a gas pressure; these formulae are general enough to be 
applied to any axisymmetric magnetic structure.  

The main goal of this paper is to perform numerical simulations of nonlinear 
torsional Alfv\'en waves in our 
developed model of magnetic flux of axial symmetry 
tube embedded in the solar atmosphere with the temperature profile given 
by Avrett \& Loeser (2008). 
We also study the energy transfer from Alfv\'en waves 
to slow magnetoacoustic waves in the non-linear regime, as well as their 
reflection and transmission through realistic solar transition region, which are 
significant in constituting vertical jet flows (mass) and localized energy 
transport processes
in the inner corona. 
The paper is organized as follows. 
Our model of a solar magnetic flux tube of axial symmetry 
embedded in the solar atmosphere is 
developed in Section 2. The results of our numerical simulations are 
presented and discussed in Section 3. Conclusions are given in Section 4.
%

%1
%%%%%%%%%%%%%%%%%%%%%%%%%%%%%%%%%%%%%%%%%%%%%%%%%%%%%%%%%%%%%%%%%%%%%%%%%%%%%%%%%%%
\begin{figure}
\centering
\includegraphics[scale=0.45,angle=0]{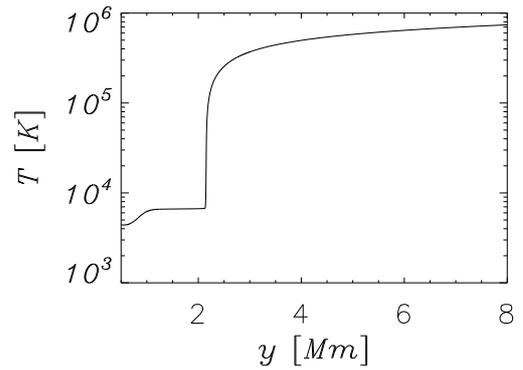}
\caption{\small
Hydrostatic equilibrium profile of solar temperature. 
}
\label{fig:initial_profile}
\end{figure}
%%%%%%%%%%%%%%%%%%%%%%%%%%%%%%%%%%%%%%%%%%%%%%%%%%%%%%%%%%%%%%%%%%%%%%%%%%%%%%%%%%
%2
%%%%%%%%%%%%%%%%%%%%%%%%%%%%%%%%%%%%%%%%%%%%%%%%%%%%%%%%%%%%%%%%%%%%%%%%%%%%%%%%%%%
\begin{figure}
\centering
\begin{center}
\includegraphics[scale=0.275, angle=0]{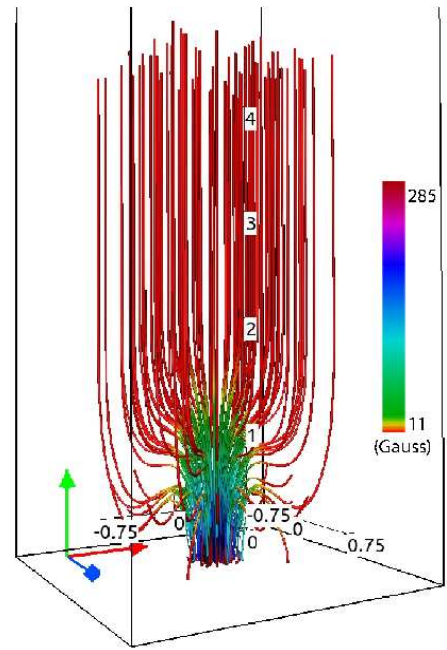}
\end{center}
\caption{\small
Equilibrium magnetic field lines. Red, green, and blue arrows 
correspond to the $x$-, $y$-, and $z$-axis, respectively. 
The size of the box shown is 
$(-0.75,0.75)\times (0,4)\times 
(-0.75,0.75)$ Mm. 
% The navy blue lines correspond to the photospheric magnetic field strength of $\sim 285$ Gauss, 
% the green lines to the chromospheric field of its strength of $\sim 55$ Gauss,
% and the red lines to the coronal field of its strength of $\sim 10$ Gauss.
The color map corresponds to the 
magnitude of a magnetic field. 
}
\label{fig:initial_B}
\end{figure}

%3
%%%%%%%%%%%%%%%%%%%%%%%%%%%%%%%%%%%%%%%%%%%%%%%%%%%%%%%%%%%%%%%%%%%%%%%%%%%%%%%%%%%%%
\begin{figure*}
\mbox{
\includegraphics[scale=0.45, angle=0]{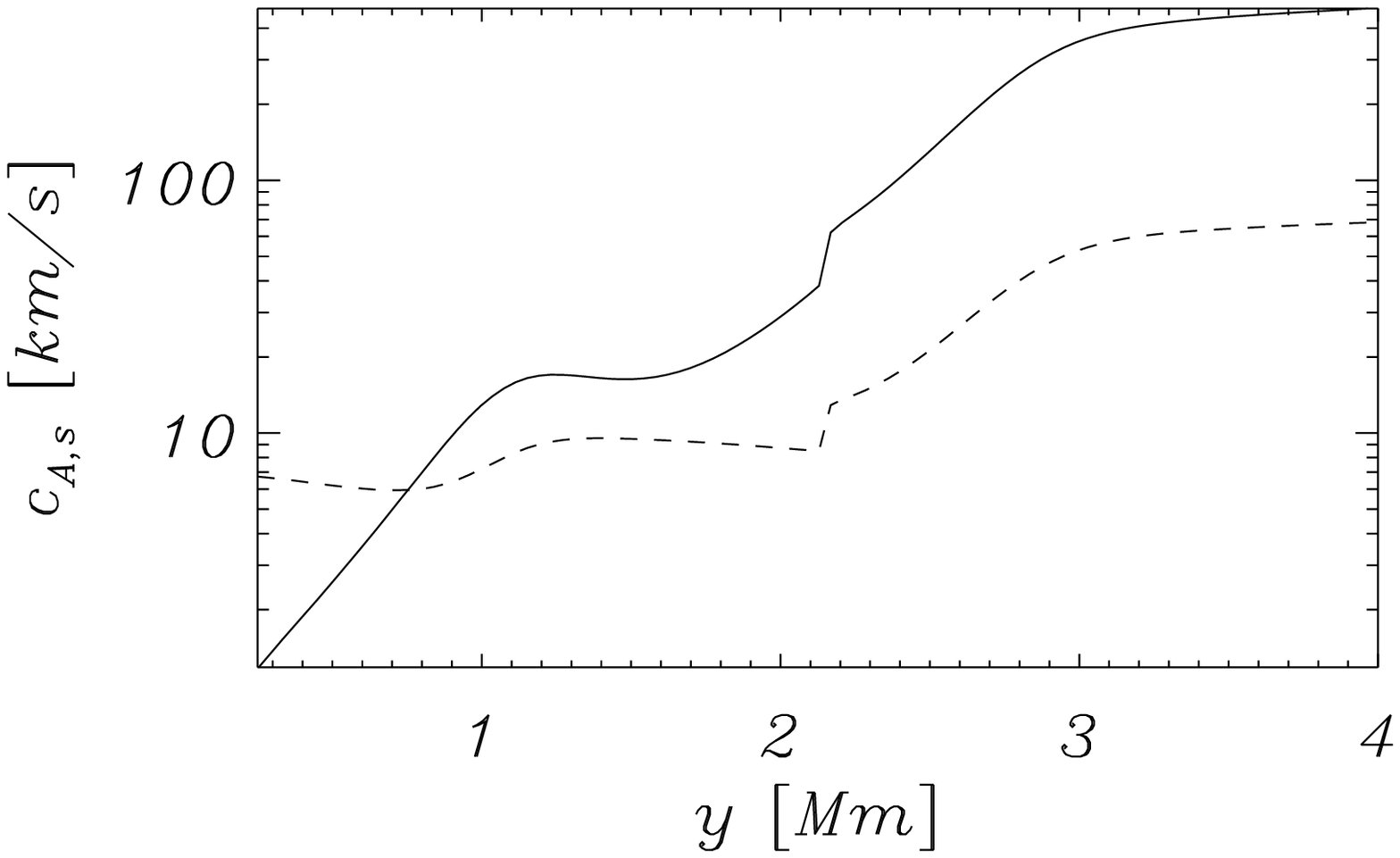}
\includegraphics[scale=0.45, angle=0]{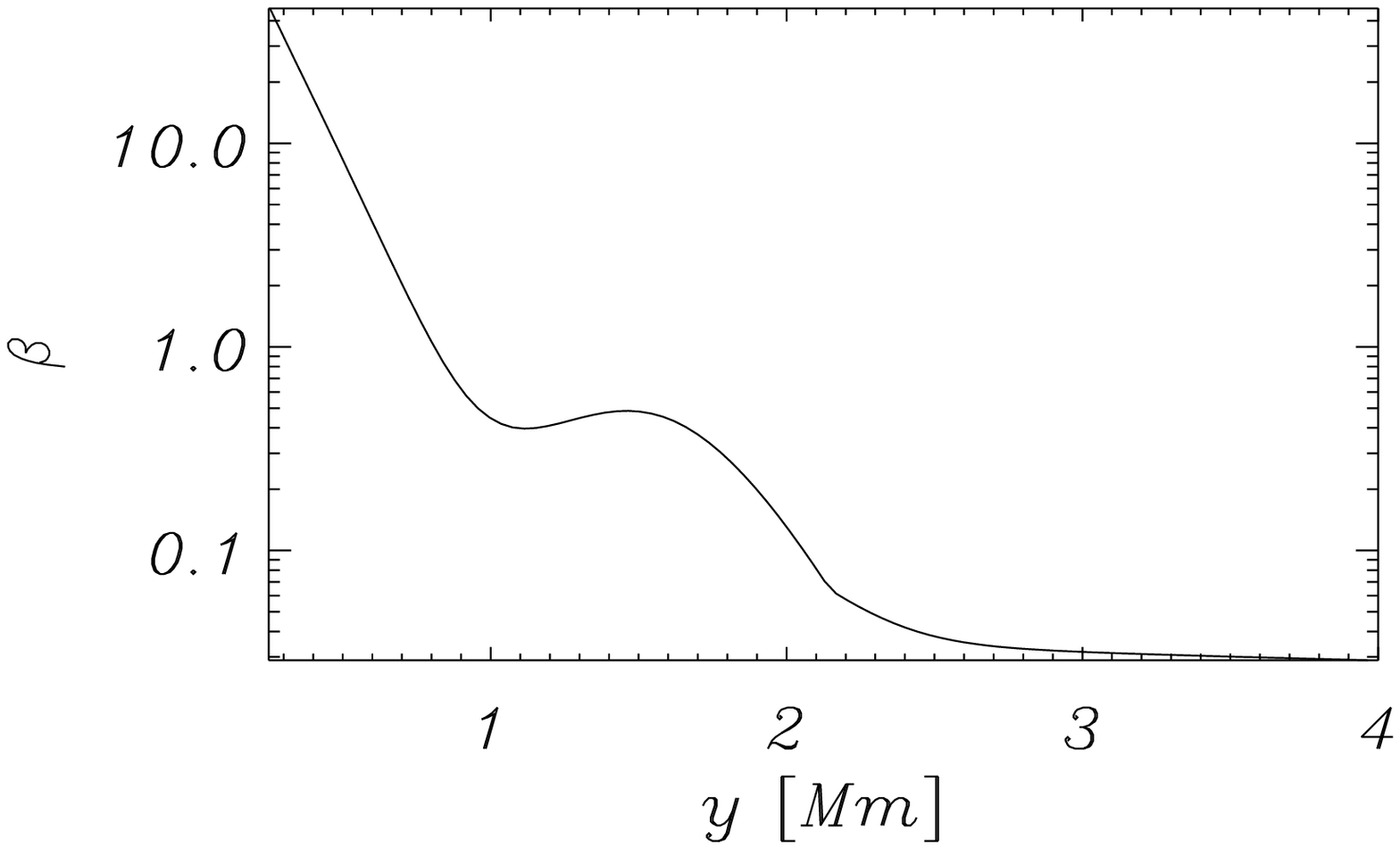}
}
\caption{\small
Equilibrium profiles of the Alfv\'en (solid) and sound (dashed) speeds (the 
left panel) and plasma $\beta$ (the right panel) along the flux tube axis. 
}
\label{fig:cs-cA}
\end{figure*}

\section{A model of a solar magnetic flux tube of axial symmetry}
We consider a model solar atmosphere that is described by the following ideal, 
adiabatic, 3D magnetohydrodynamic (MHD) equations: 
\beqa
\label{eq:MHD_rho}
{{\partial \varrho}\over {\partial t}}+\nabla \cdot (\varrho{\bf V})=0\, ,
\\
\label{eq:MHD_V}
\varrho{{\partial {\bf V}}\over {\partial t}}+ \varrho\left ({\bf V}\cdot 
\nabla\right ){\bf V} = -\nabla p+ \frac{1}{\mu}(\nabla\times{\bf B})
\times{\bf B} +\varrho{\bf g}
\, ,
\\
\label{eq:MHD_p}
{\partial p\over \partial t} + \nabla\cdot (p{\bf V}) = (1-\gamma)p 
\nabla \cdot {\bf V}\, ,
\hspace{3mm}
p = \frac{k_{\rm B}}{m} \varrho T\, ,
\\
\label{eq:MHD_B}
{{\partial {\bf B}}\over {\partial t}}= \nabla \times ({\bf V}\times{\bf B})\, ,
\hspace{3mm}
\nabla\cdot{\bf B} = 0\ ,
\eeqa
where ${\varrho}$ is mass density, ${\bf V}$ is the flow velocity, and ${\bf B}$
is the magnetic field.  The standard notation is used for the other physical 
parameters, namely, $p$, $T$, $\gamma=5/3$, $\mu$, ${\bf g}=(0,-g,0)$, $m$, 
and $k_{\rm B}$ are the gas pressure, temperature, adiabatic index, magnetic 
permeability, mean particle mass, and Boltzmann's constant, respectively.  The 
magnitude of the gravitational acceleration is $g=274$ m s$^{-2}$.  The value 
of $m$ is specified by the mean molecular weight, which is $1.24$ in the solar photosphere (Oskar Steiner, 
private communication) and assumed to be constant in the entire flux tube model.
%

%4
%%%%%%%%%%%%%%%%%%%%%%%%%%%%%%%%%%%%%%%%%%%%%%%%%%%%%%%%%%%%%%%%%%%%%%%%%%%%%%%%%%%
\begin{figure}
\centering
%\vspace{-2.5cm}
\includegraphics[scale=0.5, angle=0]{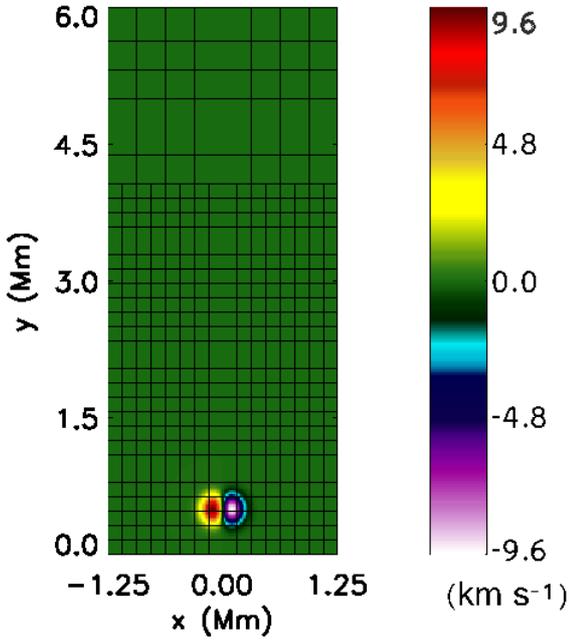}
%\vspace{-1.5cm}
\caption{\small 
Numerical blocks with their boundaries (solid lines) and the pulse in velocity 
$V_{\rm z}$ of Eq.~(\ref{eq:perturb}) (color maps) in the $x-y$ plane for $z=0$ 
Mm at $t=0$ seconds. A part of the simulation region is only displayed. 
}
\label{fig:amr}
\end{figure}
%%%%%%%%%%%%%%%%%%%%%%%%%%%%%%%%%%%%%%%%%%%%%%%%%%%%%%%%%%%%%%%%%%%%%%%%%%%%%%%%%%%

We assume that the solar atmosphere is in static equilibrium (${\bf V}_{\rm e}={\bf 0}$) 
with the Lorentz force balanced by the gravity force and the gas pressure gradient, 
which means that 
\begin{equation}
\label{eq:force_free}
\frac{1}{\mu}(\nabla\times{\bf B}_{\rm e})\times{\bf B}_{\rm e} + 
\varrho_{\rm e} {\bf g} -\nabla p_{\rm e} = {\bf 0}\ ,
\end{equation}
\noindent
where the subscript $'{\rm e}'$ corresponds to the equilibrium configuration. 
We consider an axisymmetric flux tube whose equilibrium is described by 
Eq.~(\ref{eq:force_free}) and its magnetic field satisfies the solenoidal 
condition.   We now introduce the magnetic flux ($\Psi$) normalized by 
2$\pi$, and obtain 
\beq\label{eq:psi}
\Psi(r,y) = \int_0^r B_{\rm ey} 
r'\, dr'\, ,
\eeq
where $B_{\rm ey}$ is a vertical magnetic field component and $r$ is the 
radial distance from the flux tube axis given by
\beq
r = \sqrt{x^2+z^2}\, .
\eeq

As a result of Eq.~(\ref{eq:psi}), the magnetic field is automatically 
divergence-free and its radial ($B_{\rm er}$), azimuthal ($B_{\rm 
e\theta}$) and vertical ($B_{\rm ey}$) components are expressed as 
\beq\label{eqs:Br-By}
B_{\rm er}    = -\frac{1}{r}
\frac{\partial \Psi}{\partial y}\, ,\hspace{3mm}
B_{\rm e\theta} = 0\, ,\hspace{3mm}
B_{\rm ey}    =  \frac{1}{r}
\frac{\partial \Psi}{\partial r}\, .
\eeq
We consider a magnetic flux tube of axial symmetry, 
which is initially non-twisted, and 
represent the tube magnetic field in terms of its azimuthal component of 
the vector-potential ($A{\bf\hat \theta}$) as
\beq\label{eq:equil_B}
{\bf B_{\rm e}} = \nabla\times A{\bf\hat \theta}\, ,
\eeq
where ${\bf\hat \theta}$ is a unit vector along the azimuthal direction. 
In this case, we have
\beq\label{eq:B_com_A}
B_{\rm er} = -\frac{\partial A}{\partial y}\, ,\hspace{3mm} 
B_{\rm e\theta} = 0\, , \hspace{3mm} 
B_{\rm ey} = \frac{1}{r} \frac{\partial (rA)}{\partial r}\, .
\eeq

Comparing Eqs.~(\ref{eqs:Br-By}) and (\ref{eq:B_com_A}), we find 
\beq
A(r,y) = \frac{1} {r} \Psi(r,y) \, .
\eeq
In Cartesian coordinates the magnetic vector potential components 
are 
\beq
[A_{\rm x}, A_{\rm y}, A_{\rm z}] = [A \frac{z}{r}, 0, -A \frac{x}{r}]\, .
\eeq
These formulae are useful for numerical simulations because they identically
satisfy the selenoidal condition, which is important in controlling numerical
errors. 

Multiplying now Eq.~(\ref{eq:force_free}) by ${\bf B_{\rm e}}$, we obtain
\beq
{\bf B_{\rm e}}\cdot (\nabla p_{\rm e} - \varrho_{\rm e}{\bf g}) = 0\, .
\eeq
Since $p_{\rm e}=p_{\rm e}(\Psi, y)$, therefore, from the above equation 
with the use of Eq.~(\ref{eqs:Br-By}), we have 
\beq\label{eq:hydro_stat}
\varrho_{\rm e} g = -\frac{\partial p_{\rm e}(\Psi,y)}{\partial y}\, ,
\eeq
which means that the hydrostatic condition is satisfied along magnetic field 
lines, and that $\Psi$ = const. 

Rewriting Eq.~(\ref{eq:force_free}) in its components and taking Eq.~(\ref{eqs:Br-By}) 
into account, we obtain the so-called Grad-Shafranov equation (e.g., Low 1975; Priest 
1982) given by 
\beq\label{eq:grad-shaf}
\frac{\partial^2 \Psi}{\partial r^2} - \frac{1}{r}\frac{\partial \Psi}{\partial r} 
+ \frac{\partial^2 \Psi}{\partial y^2} = - \mu r^2 \frac{\partial p_{\rm e}(\Psi,y)}
{\partial \Psi}\, .
\eeq

If we consider the equilibrium gas pressure as a given function, then 
Eq.~(\ref{eq:grad-shaf}) becomes a nonlinear Dirichlet problem for $\Psi$. 
In contrast, Low (1980) proposed the inverse problem by specifying magnetic 
field through a choice of $\Psi$ by integrating Eq.~(\ref{eq:grad-shaf})
and deriving the general formulas for $p_{\rm e}$ and $\varrho_{\rm e}$ 
given in terms of $\Psi$ and its derivatives.  In our approach, we regard
$y$ as a parameter, and after integrating Eq.~(\ref{eq:grad-shaf}) over 
$\Psi$, we find (Solov'ev 2010)
\beq\label{eq:eq_gas_press}
p_{\rm e}(r,y) = p_{\rm h}(y) - \frac{1}{\mu} 
\left[ 
\frac{1}{2r^2}\left(\frac{\partial \Psi}{\partial r}\right)^2 + 
\int \frac{\partial^2 \Psi}{\partial y^2}\frac{\partial \Psi}{\partial r}
\frac{dr}{r^2} \right]\, ,
\eeq
where 
\beq
\label{eq:pres}
p_{\rm h}(y)=p_{\rm 0}~{\rm exp}\left[ -\int_{y_{\rm r}}^{y}\frac{dy^{'}}
{\Lambda (y^{'})} \right]\, ,
\eeq
is the hydrostatic gas pressure, 
%(Murawski et al. 2013), 
and
\begin{equation}
\Lambda(y) = \frac{k_{\rm B} T_{\rm h}(y)} {mg}
\end{equation}
is the pressure scale-height and $T_{\rm h}(y)$ is a hydrostatic 
temperature profile. 

We adopt $T_{\rm h}(y)$ for the solar atmosphere that is specified by the 
model developed by Avrett \& Loeser (2008).  This temperature profile is 
smoothly extended into the corona (Fig.~\ref{fig:initial_profile}).  It 
should be noted that in our model the solar photosphere occupies the region 
$0 < y < 0.5$ Mm, the chromosphere is sandwiched between $y=0.5$ Mm and the 
transition region, which is located at $y\simeq 2.1$ Mm, and that above this
height the atmospheric layers represent the solar corona.   

According to Eq.~(\ref{eq:hydro_stat}), the equilibrium mass density can be 
calculated if ${\partial p_{\rm e}(y,\Psi)}/{\partial y}$ is known.  Moreover, 
from Eq.~(\ref{eq:eq_gas_press}), we get $p_{\rm e} = p_{\rm e}(r,y)$.  Thus,
in order to find ${\partial p_{\rm e}(y,\Psi)}/{\partial y}$, we use the 
following relations that are valid for any differentiable function ($S$):
\beqa\label{eq:A1}
\frac{\partial S(r,y)}{\partial r} &=& \frac{\partial S(\Psi,y)}{\partial \Psi} 
\frac{\partial \Psi}{\partial r}\, ,\\
\frac{\partial S(r,y)}{\partial y} &=& \frac{\partial S(\Psi,y)}{\partial y} + 
\frac{\partial S(\Psi,y)}{\partial \Psi} \frac{\partial \Psi}{\partial y} \, .
\label{eq:A1a}
\eeqa

However, it follows from Eq.~(\ref{eq:eq_gas_press}) that we need to specify 
$S(r,y)$ as
\beq
S(r,y) = \frac{1}{2r^2}\left(\frac{\partial \Psi}{\partial r}\right)^2 + 
\int \frac{\partial^2 \Psi}{\partial y^2}\frac{\partial \Psi}{\partial r}
\frac{dr}{r^2}\, . 
\eeq
We calculate first ${\partial S(r,y)}/{\partial r}$, and then use Eq.~(\ref{eq:A1}) 
to determine ${\partial S(y, \Psi)}/{\partial \Psi}$.  The result is 
\beq\label{eq:pe_o_Psi}
\frac{\partial p_{\rm e}(y, \Psi)}{\partial \Psi} = 
-\frac{1}{\mu} 
\left[ 
\frac{1}{r} \frac{\partial}{\partial r}\left( \frac{1}{r}\frac{\partial \Psi}
{\partial r} \right) + \frac{1}{r^2} \frac{\partial^2\Psi}{\partial y^2}\right]\, .
\eeq
Using Eq.~(\ref{eq:A1a}), we get
\beq
\frac{\partial p_{\rm e}(y, \Psi)}{\partial y} = 
\frac{\partial p_{\rm e}(r,y)}{\partial y} 
- 
\frac{\partial p_{\rm e}(y,\Psi)}{\partial \Psi} \frac{\partial \Psi}{\partial y} \, .
\eeq
Finally, with help of Eq.~(\ref{eq:pe_o_Psi}), we write
\beqa\nonumber
\varrho_{\rm e}(r,y) = \varrho_{\rm h}(y) + \\
\nonumber
\frac{1}{\mu g} 
\{ 
\frac{\partial}{\partial y} 
\left[
\frac{1}{2r^2}\left( \left(\frac{\partial \Psi}{\partial r}\right)^2 - 
\left(\frac{\partial \Psi}{\partial y}\right)^2 \right)
+ 
\int \frac{\partial^2 \Psi}{\partial y^2}\frac{\partial \Psi}{\partial r}
\frac{dr}{r^2} \right]
\\
-\frac{1}{r}\frac{\partial \Psi}{\partial y}  
\frac{\partial}{\partial r} 
\left(\frac{1}{r}\frac{\partial \Psi}{\partial r}
\right)
\}
\, ,
\label{eq:eq_mass_dens}
\eeqa
with
\beq
\varrho_{\rm h} (y)=\frac{p_{\rm h}(y)}{g \Lambda(y)}\, 
\eeq
being the hydrostatic mass density. 
%(Murawski et al. 2013). 

Note that in the above formulas the magnetic flux ($\Psi$) and consequently 
the magnetic flux function ($A$) are free to choose.  For a flux tube, we 
can specify them as 
\beqa\label{eq:A-Psi-flux-tube}
A(r,y) = B_{\rm 0} \exp{(-k_{\rm y}^2 y^2)} \frac{r}{1+k_{\rm r}^4r^4} + 
\frac{1}{2}B_{\rm y0} r \, , \\
\Psi(r,y) = r A(r,y)\, ,
\eeqa
where $B_{\rm y0}$ is the external magnetic field along the vertical direction, 
$k_{\rm r}$ and $k_{\rm y}$ are inverse length scales along the radial and 
vertical directions, respectively.  The magnetic field lines, which follow 
from Eqs.~(\ref{eq:B_com_A}) and (\ref{eq:A-Psi-flux-tube}) are displayed in 
Fig.~\ref{fig:initial_B} for $k_{\rm r}=k_{\rm y}=4$ Mm$^{-1}$.  We choose 
the magnitude of the reference magnetic field $B_{\rm 0}$ in such way that 
the magnetic field within the flux tube, at $(x=0,y=0.5,z=0)$ Mm, is about 
$285$ Gauss, and $B_{\rm y0}\approx 11.4$ Gauss.  For these values, the 
resulting magnetic field lines are predominantly vertical around the line, 
$x=z=0$ Mm, while further out they are bent and ${\bf B}_{\rm e}$ decays with 
distance from this line.  It must be also noted that the magnetic field at the 
top of the simulation region is essentially uniform, with its value of $B_{\rm y0}$. 

Figure~\ref{fig:cs-cA} illustrates the sound speed, $c_{\rm s}$, (the dashed 
line in the left panel), Alfv\'en speed, $c_{\rm A}$, (the solid line in the 
same panel), and plasma $\beta$ (in the right panel) given by 
\beqa
c_{\rm s}(r,y) = \sqrt{\frac{\gamma p_{\rm e}(r,y)}{\varrho_{\rm e}(r,y)}}\, ,
\hspace{3mm}
c_{\rm A}(r,y) = \sqrt{\frac{B_{\rm e}^2(r,y)}{\mu \varrho_{\rm e}(r,y)}}\, ,\\
\beta(r,y) = \frac{2\mu p_{\rm e}(r,y)}{B_{\rm e}^2(r,y)}\, .
\eeqa
The sound speed below the transition region is smaller than $10$ km s$^{-1}$, 
however, it abruptly raises in the solar transition region, and then reaches 
its coronal value of 100 km s$^{-1}$ at the height $y=10$ Mm (not shown in the
figure).  The Alfv\'en speed reveals similar trends, reaching a value of more 
than $700$ km s$^{-1}$ at $y=10$ Mm (also not shown).  The coronal plasma $\beta$ 
is $\approx 3 \times 10^{-2}$ at $y=4$ Mm (the right panel).  It increases with the depth but remains 
lower than $1$ above the height  $y\approx 0.8$~Mm.  It should be noted that as 
a result of strong magnetic field, $\beta$ is lower within the flux tube than in 
the surrounding medium (not shown). 
\section{Results of numerical simulations}\label{sect:num_res}
Equations (\ref{eq:MHD_rho})-(\ref{eq:MHD_B}) are solved numerically using the 
code FLASH (Lee \& Deane 2009; Lee 2013).  This code implements a third-order 
unsplit Godunov solver (Murawski \& Lee 2011) with various slope limiters and 
Riemann solvers, as well as adaptive mesh refinement.  We choose the van Leer 
slope limiter and the Roe Riemann solver (Murawski \& Lee 2012).  For all the 
considered cases, we set the simulation box as $(-1.25,1.25)\, {\rm Mm} \times 
(0,20)\times (-1.25,1.25)\, {\rm Mm}$, and impose boundary conditions by fixing 
in time all plasma quantities at all six boundary surfaces to their equilibrium 
values.  In the present numerical studies, we use a non-uniform grid with a 
minimum (maximum) level of refinement set to $2$ ($5$).  The grid system at 
$t=0$ s is shown in Fig.~\ref{fig:amr} with the blocks being displayed only 
up to $y=6$ Mm.  Above this altitude the block system remains homogeneous. 
As each block consists of $8\times 8\times 8$ identical numerical cells, we 
reach the effective finest spatial resolution of $19.53$ km, below the altitude 
$y=4.25$ Mm. 

We initially perturb the azimuthal component of velocity, $V_{\theta}$, by using
\beq\label{eq:perturb}
V_{\theta}{\bf \hat \theta} = [V_{\rm x}, V_{\rm z}] = [\frac{z}{w},-\frac{x}{w}]\ A_{\rm v} 
\exp\left[ -\frac{r^2 + (y-y_{\rm 0})^2}{w^2} \right],
\eeq
where $A_{\rm v}$ denotes the amplitude of the pulse, ${\bf\hat \theta}$ is the 
unit vector along the azimuthal direction, $y_{\rm 0}$ is its vertical position, 
and $w$ is the width. We set $A_{\rm v}=150$ km s$^{-1}$, $w=150$ km, and 
$y_{\rm 0}=500$ km, and hold them fixed.  This value of $A_{\rm v}$ results in 
the effective maximum velocity of about $9.6$ km s$^{-1}$ (Fig.~\ref{fig:amr}).

%5 
%%%%%%%%%%%%%%%%%%%%%%%%%%%%%%%%%%%%%%%%%%%%%%%%%%%%%%%%%%%%%%%%%%%%%%%%%%%%%%%%%%%%%
\begin{figure*}
%{!h}
\begin{center}
\mbox{
\includegraphics[width=8.5cm,height=8.0cm, angle=0]{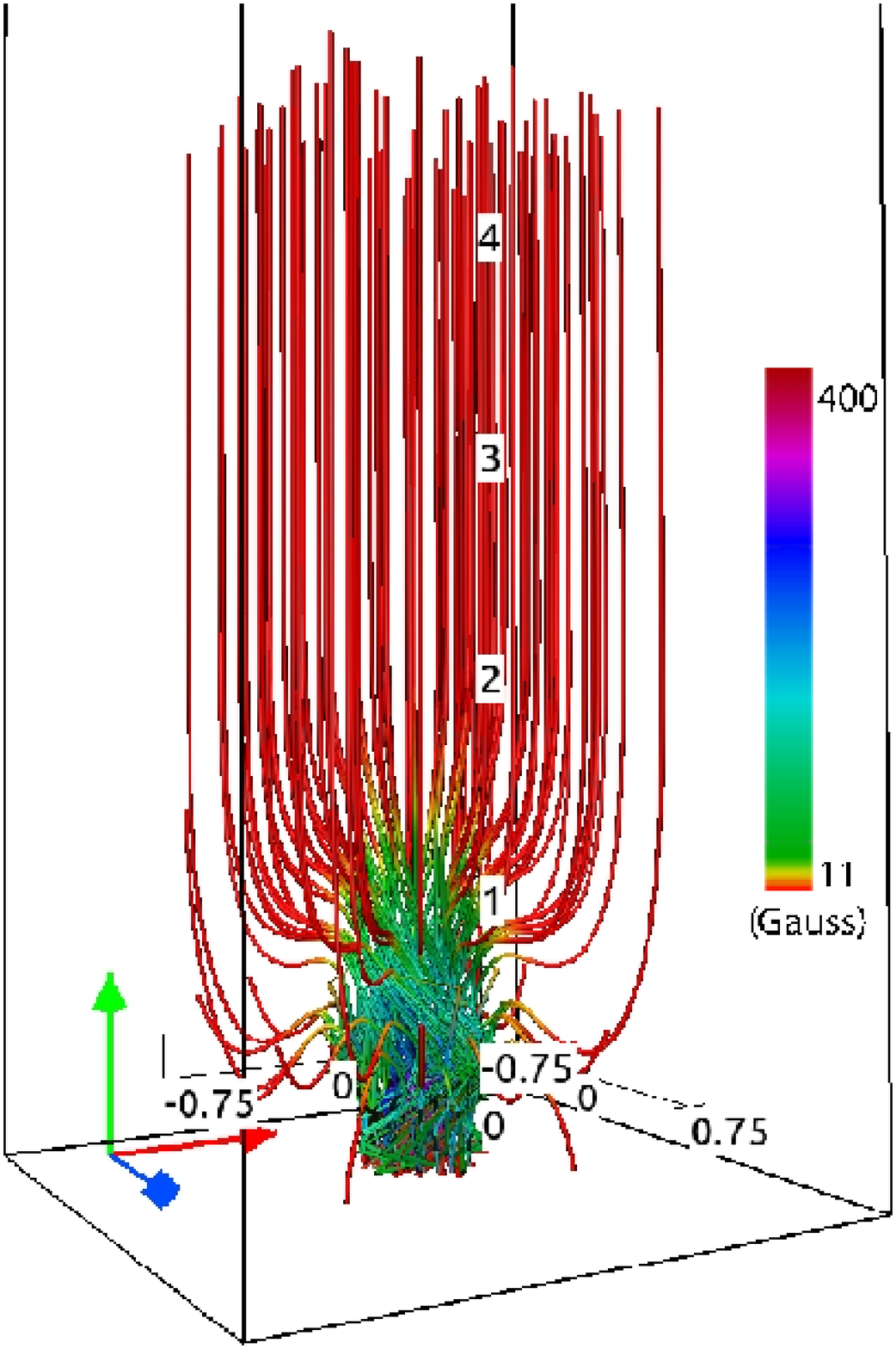}
\includegraphics[width=8.5cm,height=8.0cm, angle=0]{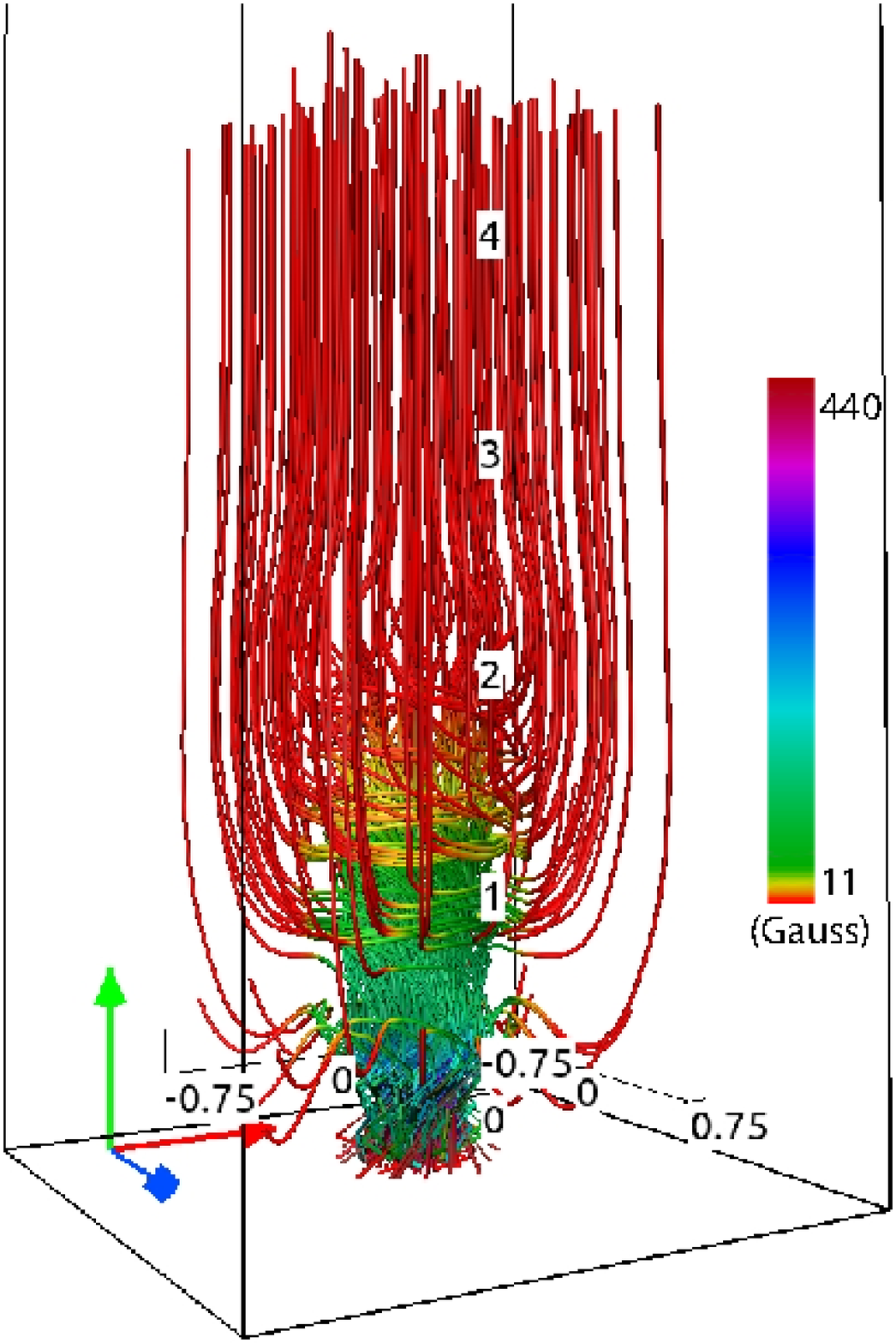}
}
\mbox{
\includegraphics[width=8.5cm,height=8.0cm, angle=0]{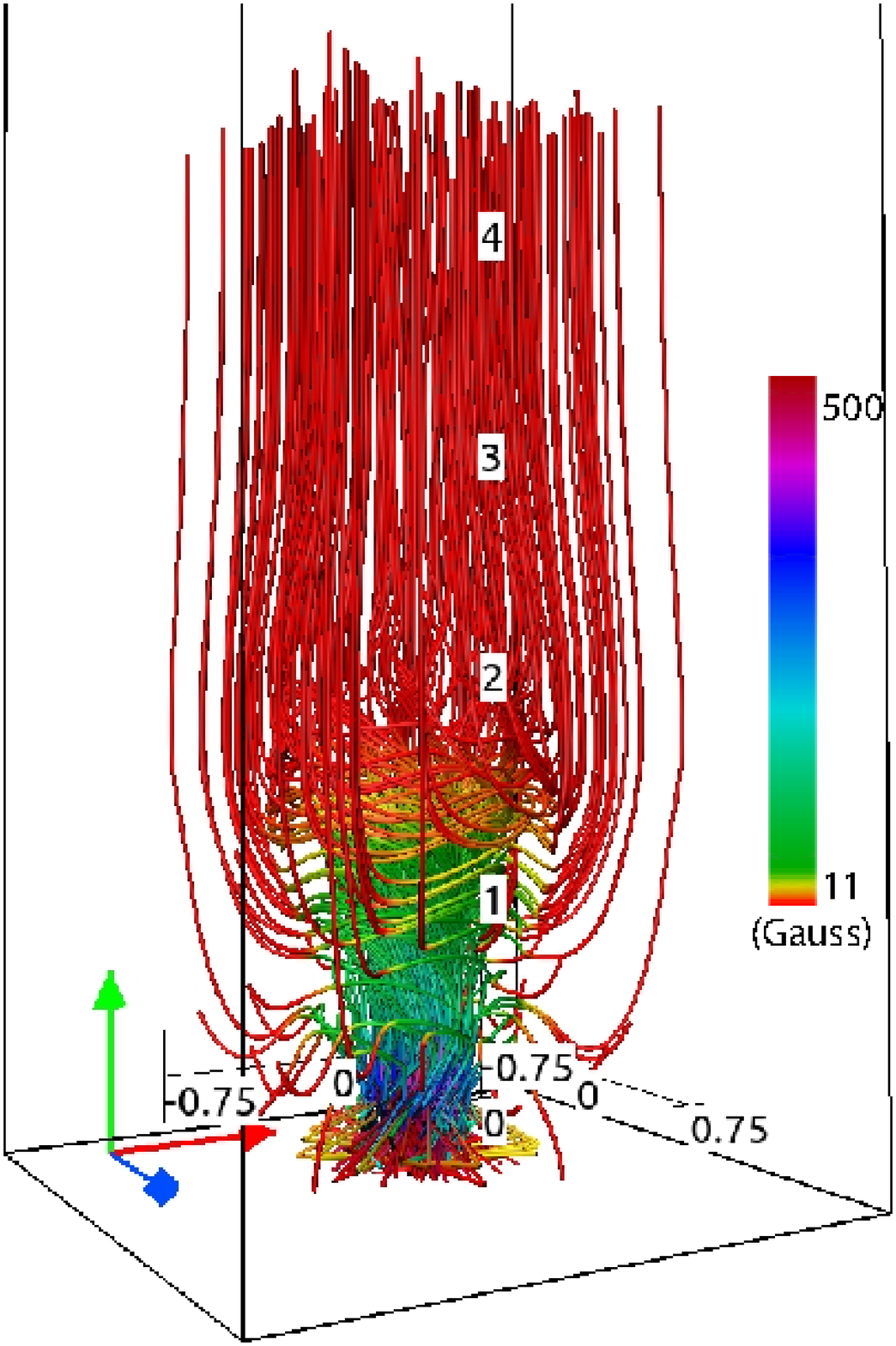}
\includegraphics[width=8.5cm,height=8.0cm, angle=0]{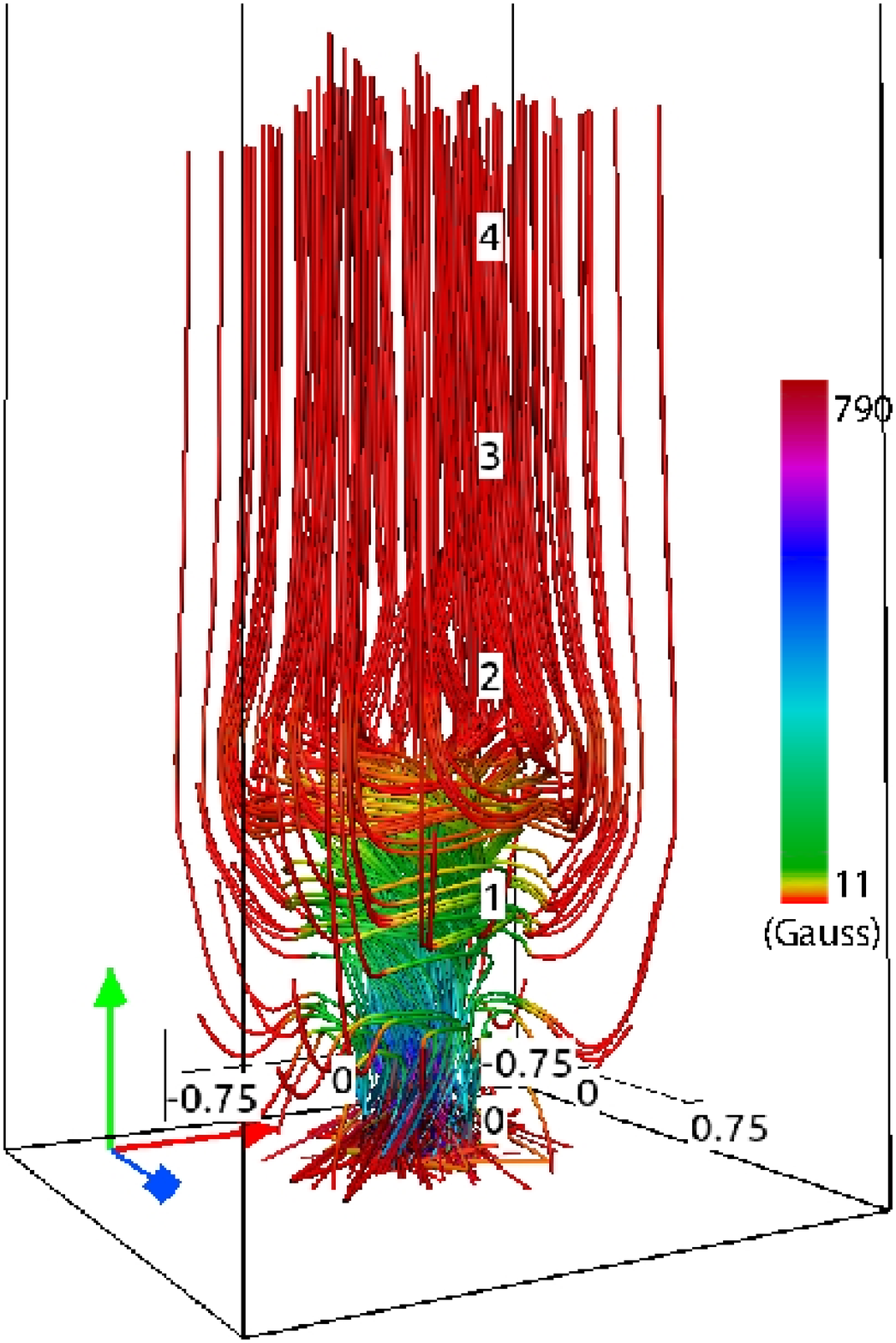}
}
\end{center}
\caption{\small
Magnetic field lines at $t=50$ s (top-left), $t=150$ s (top-right),
$t=200$ s (bottom-left), and $t=250$ s (bottom-right). Red, green and 
blue arrows design the directions of the $x$-, $y$-, and $z$-axis, 
respectively. 
The size of the box shown is 
$(-0.75,0.75)\times (0,4)\times 
(-0.75,0.75)$ Mm. 
% The navy blue lines correspond to the photospheric magnetic field strength of $\sim 285$ Gauss, 
% the green lines to the chromospheric field of its strength of $\sim 55$ Gauss,
% and 
% the red lines to the coronal field of its strength of $\sim 10$ Gauss.
The color map corresponds to the 
magnitude of a magnetic field. 
}
\label{fig:B}
\end{figure*}
%
%6
%%%%%%%%%%%%%%%%%%%%%%%%%%%%%%%%%%%%%%%%%%%%%%%%%%%%%%%%%%%%%%%%%%%%%%%%%%%%%%%%%%%%%%%%%
%%%%%%%%%%%%%%%%%%%%%%%%%%%%%%%%%%%%%%%%%%%%%%%%%%%%%%%%%%%%%%%%%%%%%%%%%%%%%%%%%%%%%%%%%%
\begin{figure*}
%{!h}
\begin{center}
%\vspace{-2.0cm}
\mbox{
\includegraphics[width=8.75cm,height=9.0cm, angle=0]{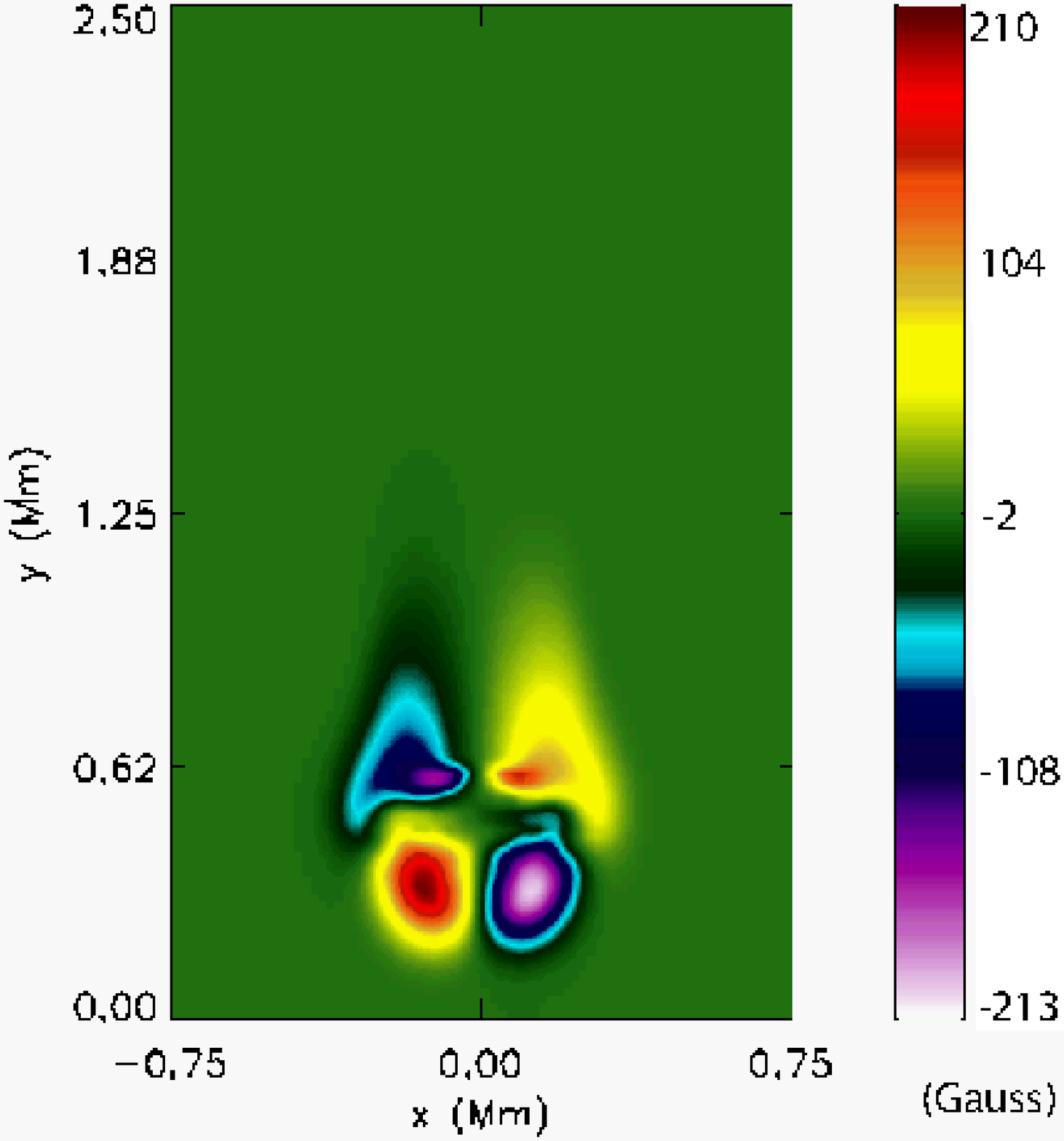}
% \hspace{-0.5cm}
  \includegraphics[width=8.75cm,height=9.0cm, angle=0]{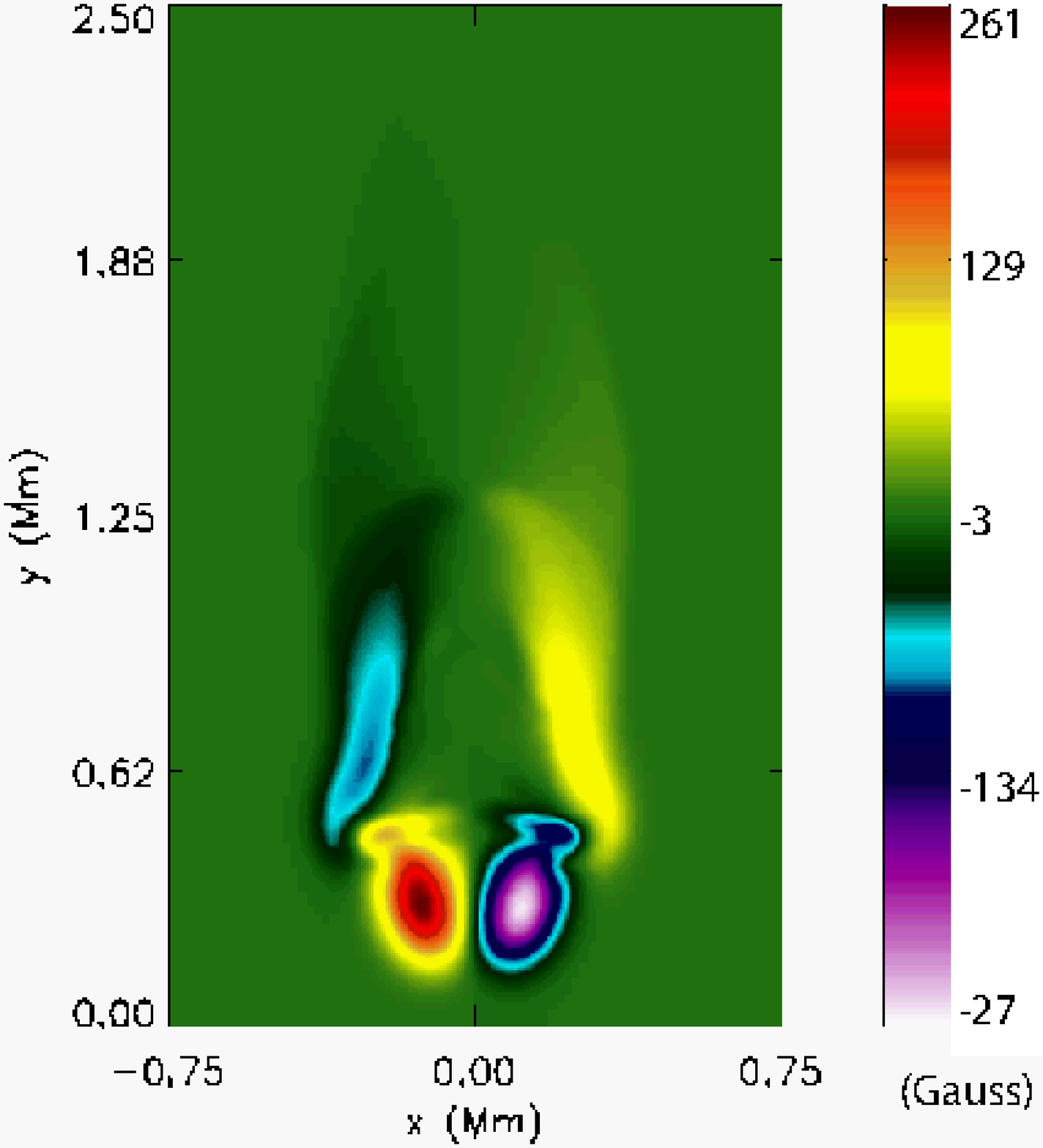}
}
\end{center}
\begin{center}
%\vspace{-2.5cm}
  \mbox{
  \includegraphics[width=8.75cm,height=9.0cm, angle=0]{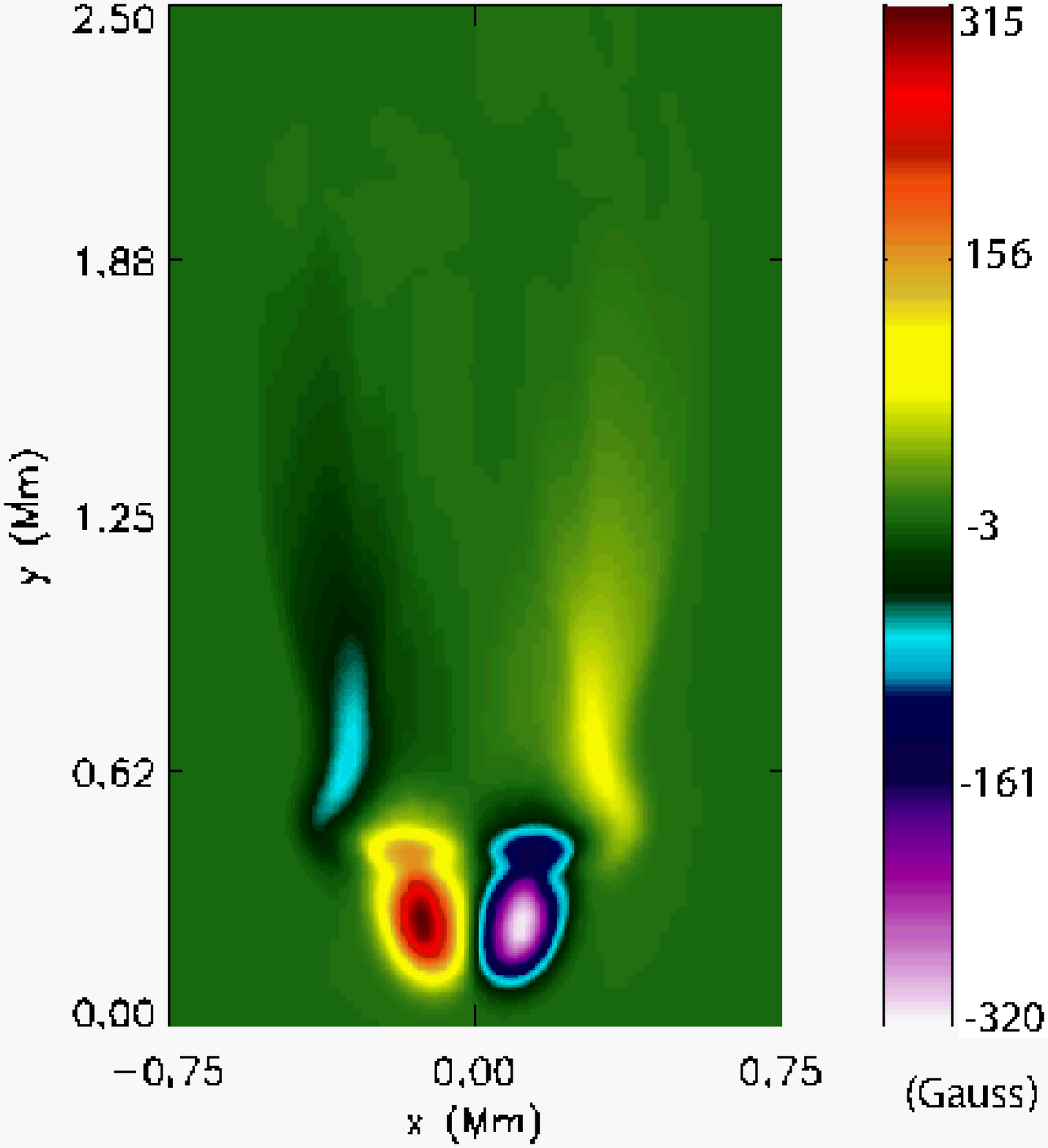}
% \hspace{-0.5cm}
  \includegraphics[width=8.75cm,height=9.0cm, angle=0]{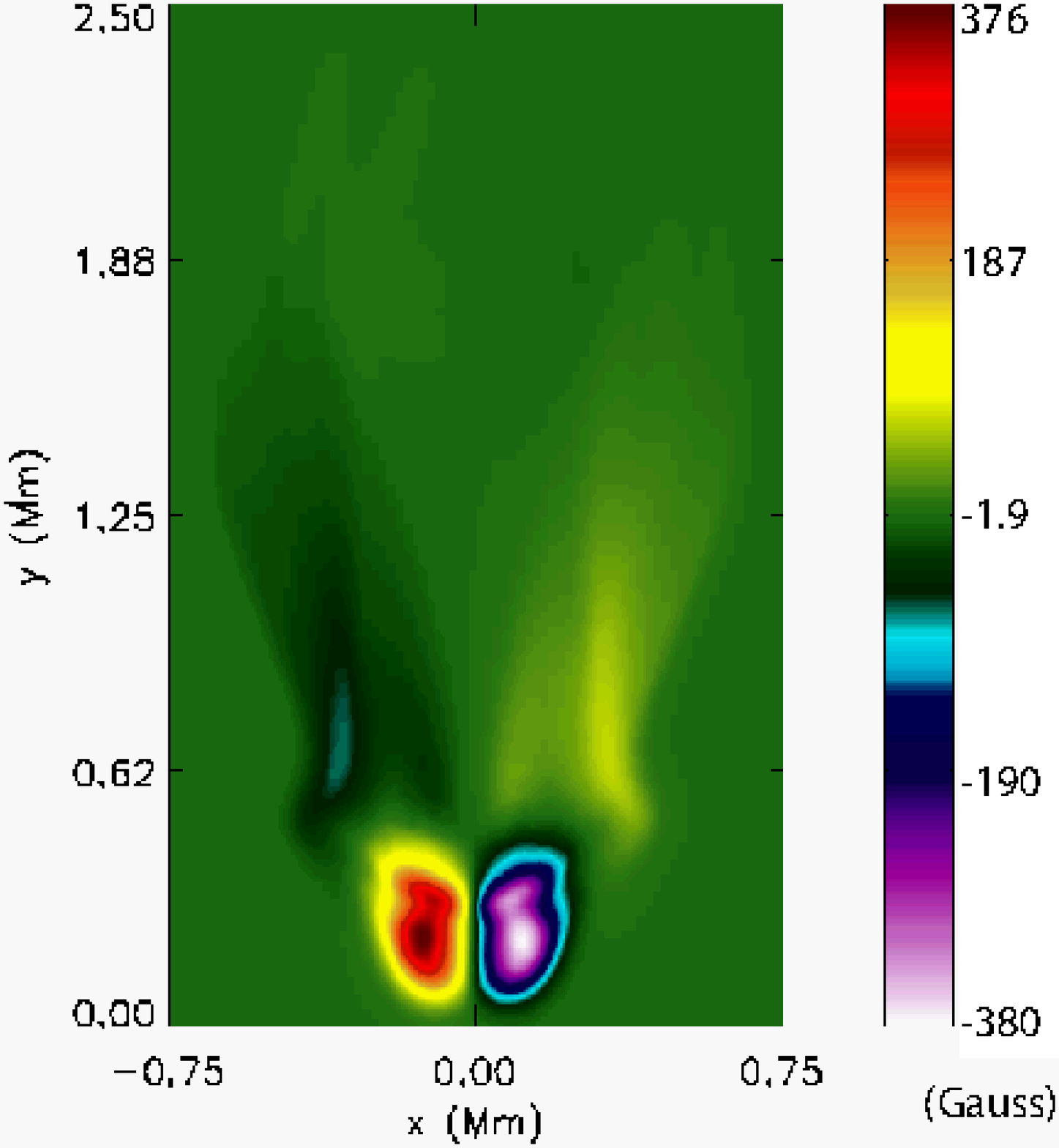}
}
\end{center}
%\vspace{-0.75cm}
\caption{\small
Spatial profiles of $B_{\rm z}(x,y,z=0)$ associated with torsional Alfv\'en 
waves, at $t=50$ s (top-left), $t=100$ s (top-right), $t=150$ s (bottom-left), 
and $t=200$ s (bottom-right). A part of the simulation region is displayed only. 
}
\label{fig:Bz}
\end{figure*}
%%%%%%%%%%%%%%%%%%%%%%%%%%%%%%%%%%%%%%%%%%%%%%%%%%%%%%%%%%%%%%%%%%%%%%%%%%%%%%%%%%%%%%%%%

The initial velocity pulse of Eq.~(\ref{eq:perturb}) triggers torsional Alfv\'en waves.  
These waves are illustrated in Fig.~\ref{fig:B}, which shows magnetic field lines at 
$t=50$ s (the top-left panel), $t=150$ s (the top-right panel), $t=200$ s (the bottom-left
panel), and $t=250$ s (the bottom-right panel).  At $t=50$ s (the top-left panel) the 
upwardly propagating Alfv\'en waves, while observed in $B_{\theta}$, reach the altitude 
$y\approx 1$ Mm, and at a later 
time Alfv\'en waves spread in space while propagating along diverged magnetic field 
lines (the top-right and bottom panels). 

Note that small-amplitude magnetic field and velocity 
perturbations that include the effect of
magnetic field expansion,
%, $B_{\theta}$, 
are governed by the following wave equations 
(Eqs. (3) \& (4) in Hollweg 1981):
\beqa
\label{eq:V_theta}
\frac{\partial^2 \left( r^2 v \right) }{\partial t^2}  = 
\frac{B_{\rm e}}{\mu \varrho_{\rm e}} \frac{\partial }{\partial s} 
\left[ r^2 B_e \frac{\partial v}{\partial s} \right]\, ,\\
\frac{\partial^2 b}{\partial t^2} = r^2 B_{\rm e} 
\frac{\partial }{\partial s} 
\left( 
\frac{B_{\rm e}}{\mu \varrho_{\rm e} r^2} 
\frac{\partial b}{\partial s} 
\right)\, ,
\label{eq:B_theta}
\eeqa
where $v=V_{\theta}/r$, $b=r B_{\theta}$, 
$V_{\theta}$ and $B_{\theta}$ are the azimuthal components of respectively perturbed velocity 
and magnetic field, and $s$ is the coordinate along the magnetic field lines. 

Moran (2001) showed that $V_{\theta}\propto \varrho_{\rm e}^{-1/4}/A B_{\rm e}$, 
where $A$ is the cross-sectional area of the flux tube. 
Near the flux tube axis we can take $A B_{\rm e}$ approximately constant, hence 
in that region $V_{\theta}\propto \varrho_{\rm e}^{-1/4}$. 
On the other hand, $B_{\theta}\propto \varrho_{\rm e}^{1/4}$ (Moran 2001).
Hence, near the flux tube axis, these
quantities are only dependent on the variation of $\varrho_{\rm e}$, 
with $V_{\theta}$ and $B_{\theta}$ respectively growing and decreasing with height.
As a result of that 
the perturbations of magnetic field lines remain weak in the solar corona 
(see also Murawski \& Musielak 2010 for the corresponding analysis 
in the case of the straight vertical magnetic field lines). 

\begin{figure}
\begin{center}
% \vspace{-1.5cm}
 \includegraphics[scale=0.5, angle=0]{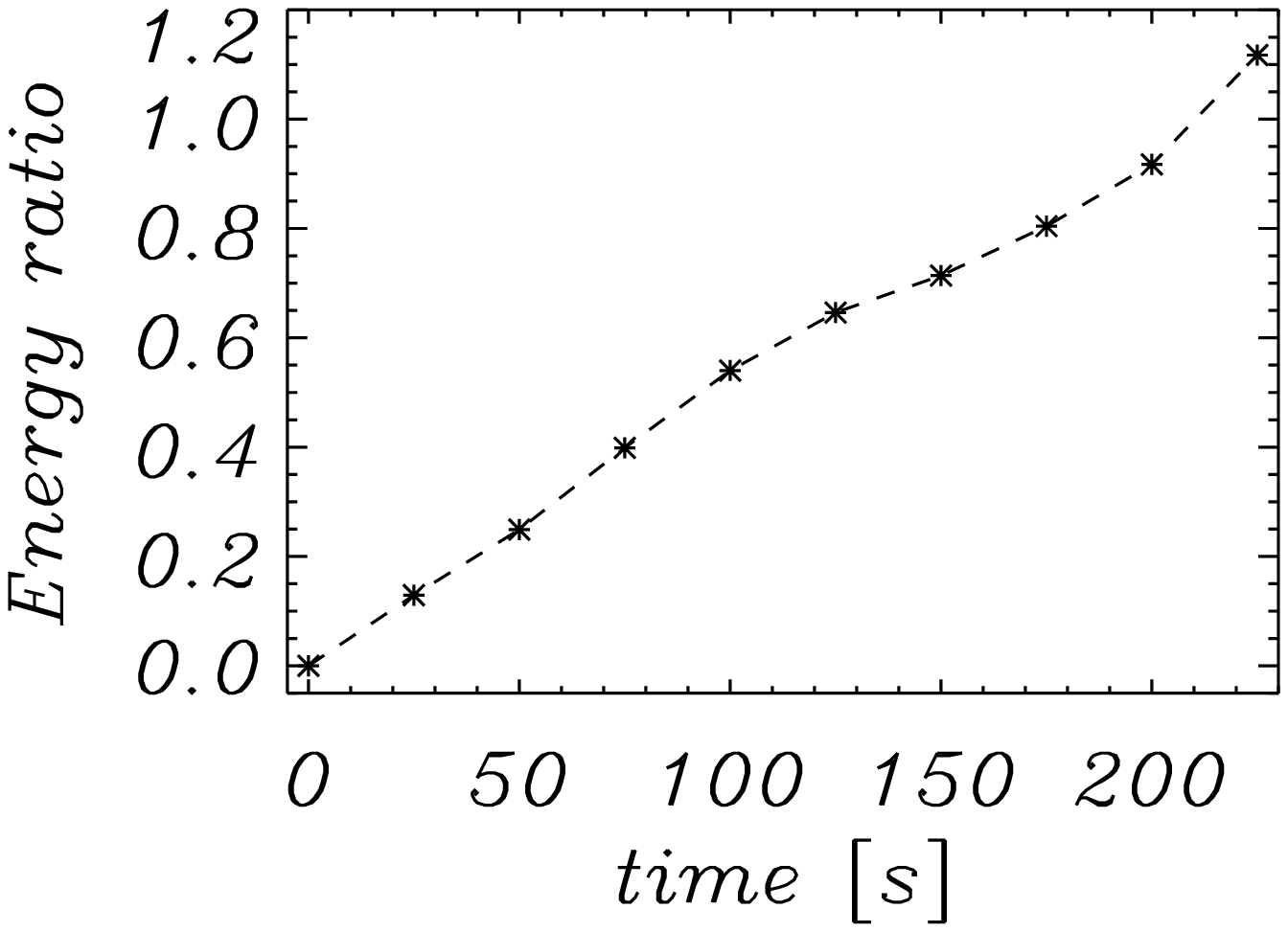}
 \includegraphics[scale=0.5, angle=0]{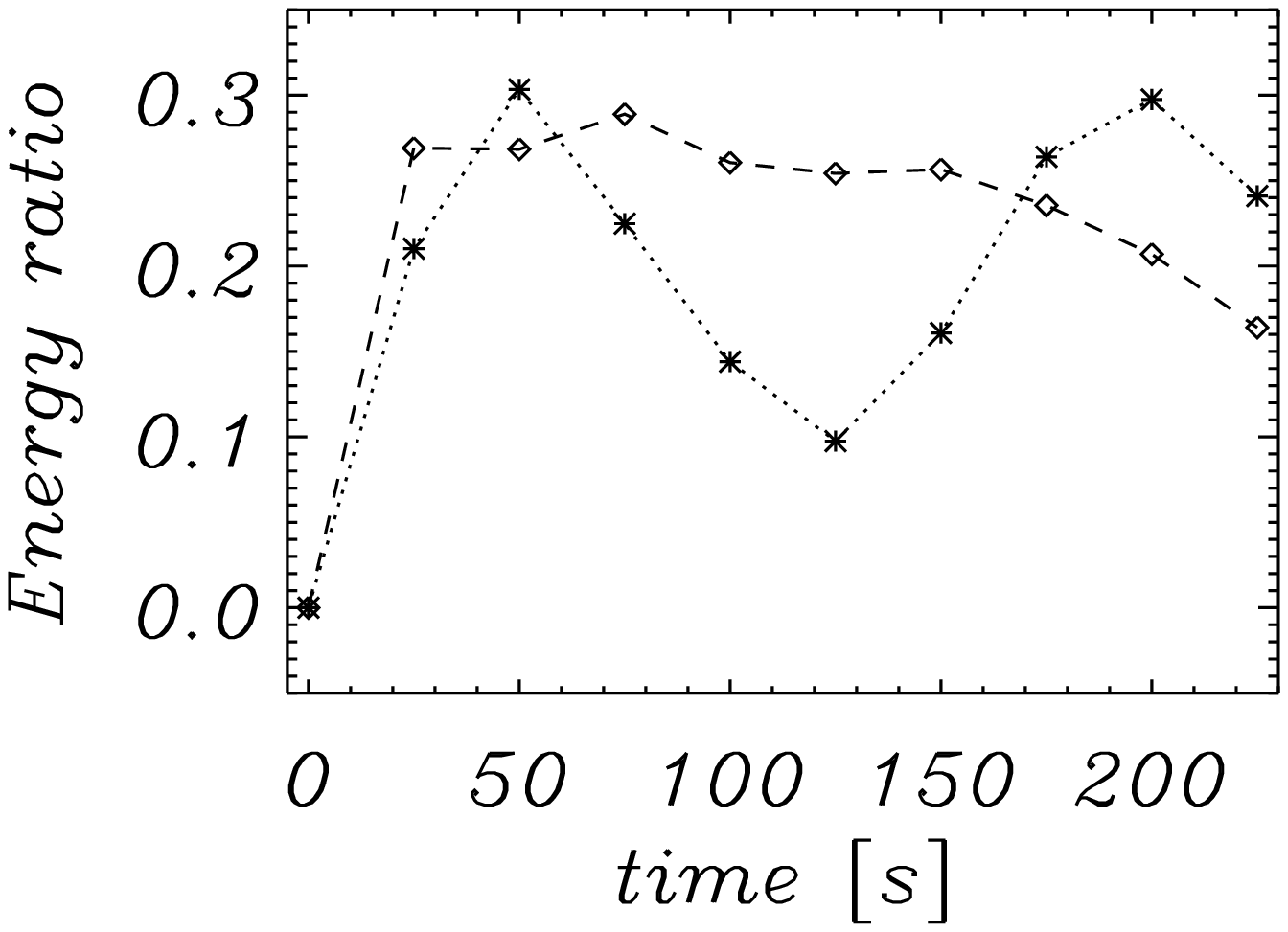}
% \vspace{-1.5cm}
\end{center}
\caption{\small 
The top panel: ratio of magnetic to kinetic energies of Alfv\'en waves; 
the bottom panel: ratio of slow magnetoacoustic waves kinetic energy to Alfv\'en waves energy 
(asterisks) and ratio of fast magnetoacoustic waves kinetic energy to total energy 
of Alfv\'en waves (diamonds). 
These energies are evaluated by integration in space over the entire computational domain. 
}
\label{fig:ener}
\end{figure}

Ratio of magnetic to kinetic energies 
(evaluated in the entire computational domain) 
of Alfv\'en waves is shown in Fig.~\ref{fig:ener} (top). 
Initially, at $t=0$ s, as we launch the initial pulse in the azimuthal velocity alone 
%Hence 
the magnetic energy is zero at $t = 0$ s but, 
according to our numerical simulations, contributions of the magnetic energy grow in time. 
It is well known that for torsional Alfv\'en waves, there is the equipartition between 
potential and kinetic energies. 
Hence according to the top panel of Fig.~\ref{fig:ener}, this equipartition of energy is not reached 
in the numerical simulation until about $t = 200$ s (where the energy ratio equals unity). 
This is the natural explanation for the limiting value of the energy ratio being approximately unity 
at $t = 200$ s. 
At $t=225$ s the magnetic energy is higher than the kinetic energy with the ratio equal to $\approx 1.1$.
Ratios of slow magnetoacoustic waves energy to Alfv\'en waves energy (asterisks) and fast magnetoacoustic waves energy to 
Alfv\'en waves energy (diamonds) is illustrated in Fig.~\ref{fig:ener} (bottom). 
At $t=0$~s kinetic energies of both the slow and fast magnetoacoustic waves are zero but as these waves are driven by 
the ponderomotive force which results from Alfv\'en waves the energy ratio grows in time reaching a maximum level of about $0.25$ 
which means that about $25\%$ of the Alfv\'en wave energy was converted into kinetic energies of magnetoacoustic waves. 
These waves propagate along magnetic field lines and power the vertical plasma flows in form of the jets.

%7
%%%%%%%%%%%%%%%%%%%%%%%%%%%%%%%%%%%%%%%%%%%%%%%%%%%%%%%%%%%%%%%%%%%%%%%%%%%%%%%%%%%%%%%%%%
%%%%%%%%%%%%%%%%%%%%%%%%%%%%%%%%%%%%%%%%%%%%%%%%%%%%%%%%%%%%%%%%%%%%%%%%%%%%%%%%%%%%%%%%%
% 
%8
%%%%%%%%%%%%%%%%%%%%%%%%%%%%%%%%%%%%%%%%%%%%%%%%%%%%%%%%%%%%%%%%%%%%%%%%%%%%%%%%%%%
\begin{figure*}
%{!h}
\centering
\includegraphics[scale=0.45,angle=0]{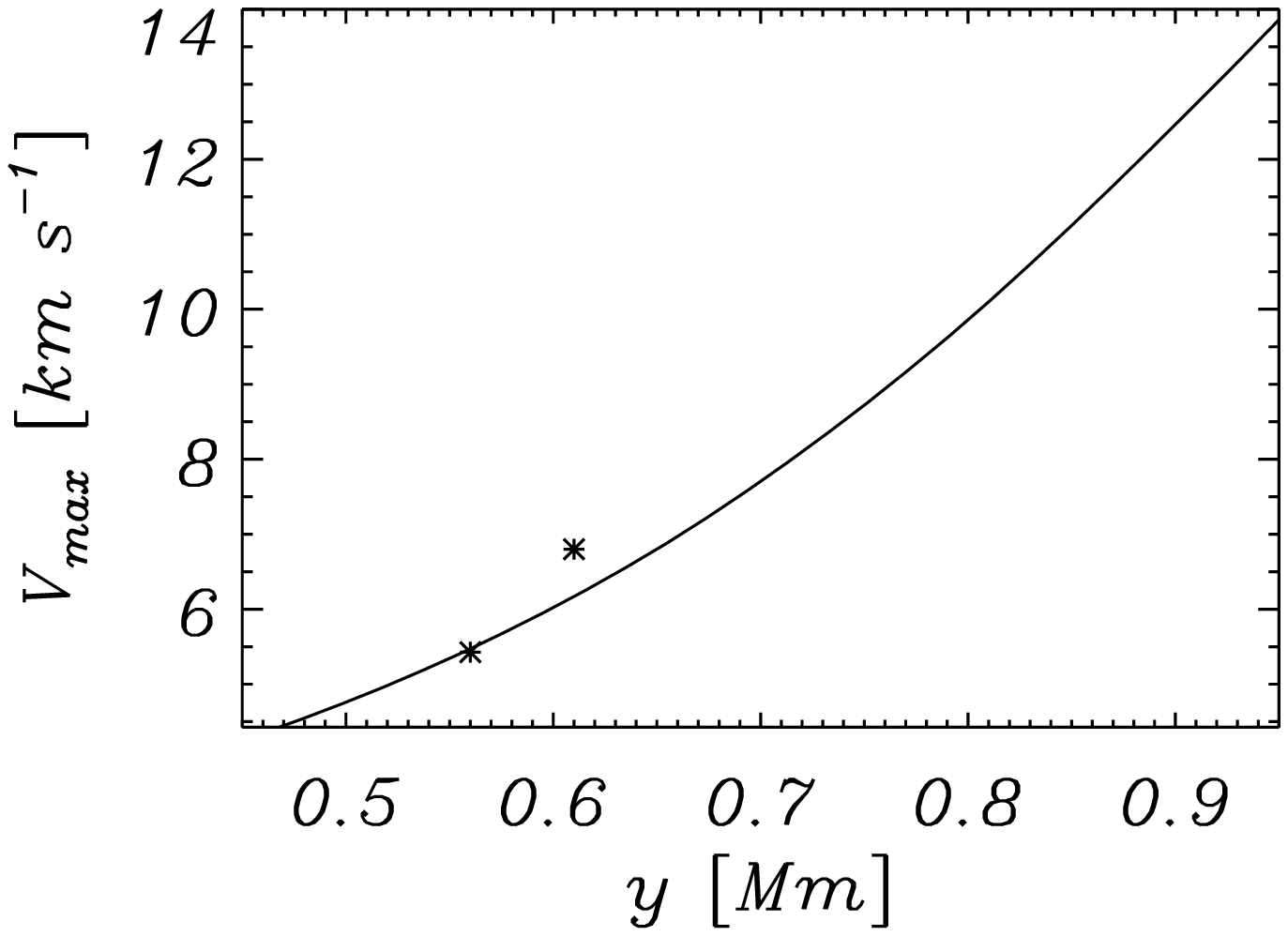}
\includegraphics[scale=0.45,angle=0]{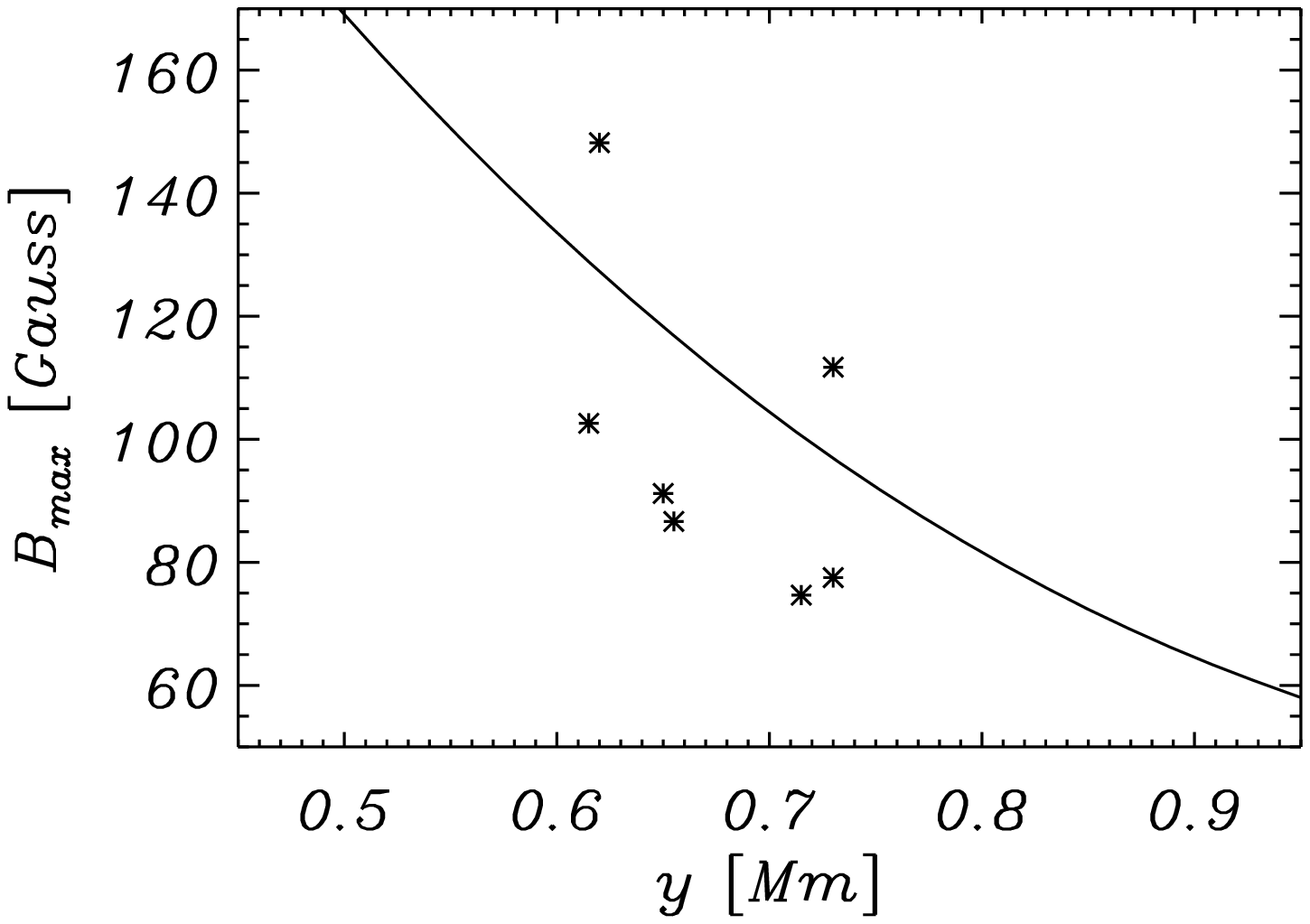}
\caption{\small
Maximum of $V_{\theta}$ (left-panel) and $B_{\theta}$ (right-panel) vs height. 
The numerical (analytical) data points are represented by asterisks (solid line).
}
\label{fig:BV-theta-vs-y}
\end{figure*}
%%%%%%%%%%%%%%%%%%%%%%%%%%%%%%%%%%%%%%%%%%%%%%%%%%%%%%%%%%%%%%%%%%%%%%%%%%%%%%%%%%
%
Figure~\ref{fig:BV-theta-vs-y} illustrates maximum of $V_{\theta}$ (left-panel) and $B_{\theta}$ (right-panel) 
as evaluated from the numerical data. 
We clearly see that 
$V_{\theta}$ grows with height 
%but the growth is much less rapid 
%than would be 
as 
predicted on the basis of the liner theory approximation reported by Moran (2001). 
On the other hand, 
$B_{\theta}$ exhibits a fall-off with $y$ with the numerical data points being close to 
the linear theory data (solid-line) as reported by Moran (2001). 
It must also be noted that the relationships shown by Moran (2001) 
are only valid when there is the equipartition of energy for torsional Alfv\'en waves. 
According to the top panel of Fig.~\ref{fig:ener}, this occurs only the time of our 
numerical simulations is $t = 200$ s or longer.  As the results of Fig.~\ref{fig:BV-theta-vs-y} 
are given for the times shorter than $t = 200$ s, then comparison of these results with the 
analytical findings of Moran (2001) is not strictly valid. 
This fall-off is also clearly evident in the spatial profiles of $B_{\rm z}(x,y,z=0)$ as 
displayed at $t=150$ s and $t=200$ s (Fig.~\ref{fig:Bz}, bottom panels).  At 
$t=50$ s and $t=100$ s (top panels), the perturbations of $B_{\rm z}(x,y,z=0)$ 
are localized below the transition region but at later times the perturbations 
that penetrate the inner solar corona are of significantly lower magnitude than those remaining 
below the transition region.  Consequently, these profiles demonstrate clearly that 
Alfv\'en waves, while traced in perturbed magnetic field lines, are essentially 
located below the transition region, which agrees with the conclusions of Murawski 
\& Musielak (2010) who used the analytical and numerical techniques for the 1D wave 
equations and found that magnetic field perturbations, while excited below the 
transition region, remain week in the solar corona in comparison to the perturbations 
in lower atmospheric layers. 

\begin{figure*}
%\centering
\begin{center}
\mbox{
\includegraphics[width=8.5cm,height=9.0cm, angle=0]{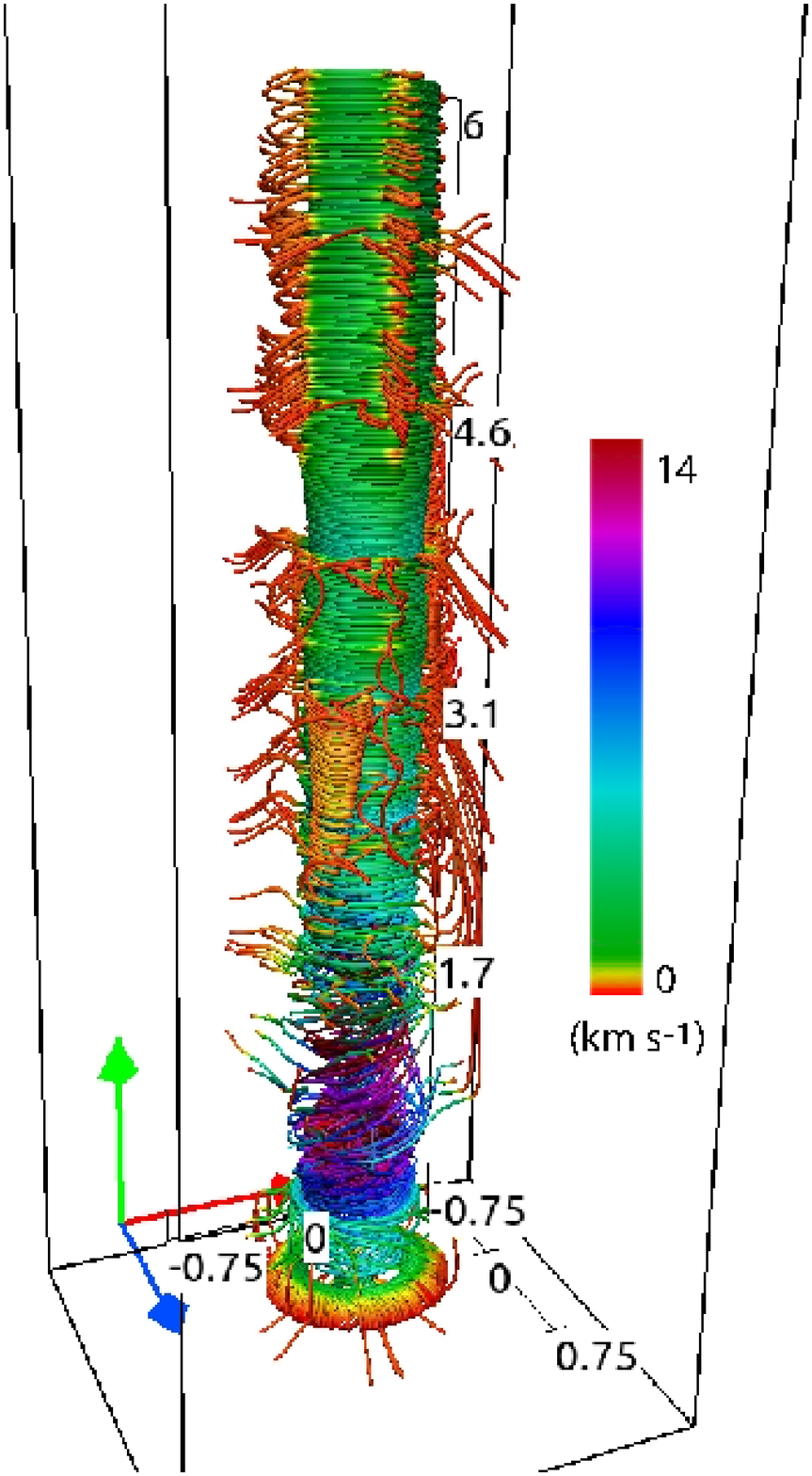}
%\hspace{-1.0cm}
\includegraphics[width=8.5cm,height=9.0cm, angle=0]{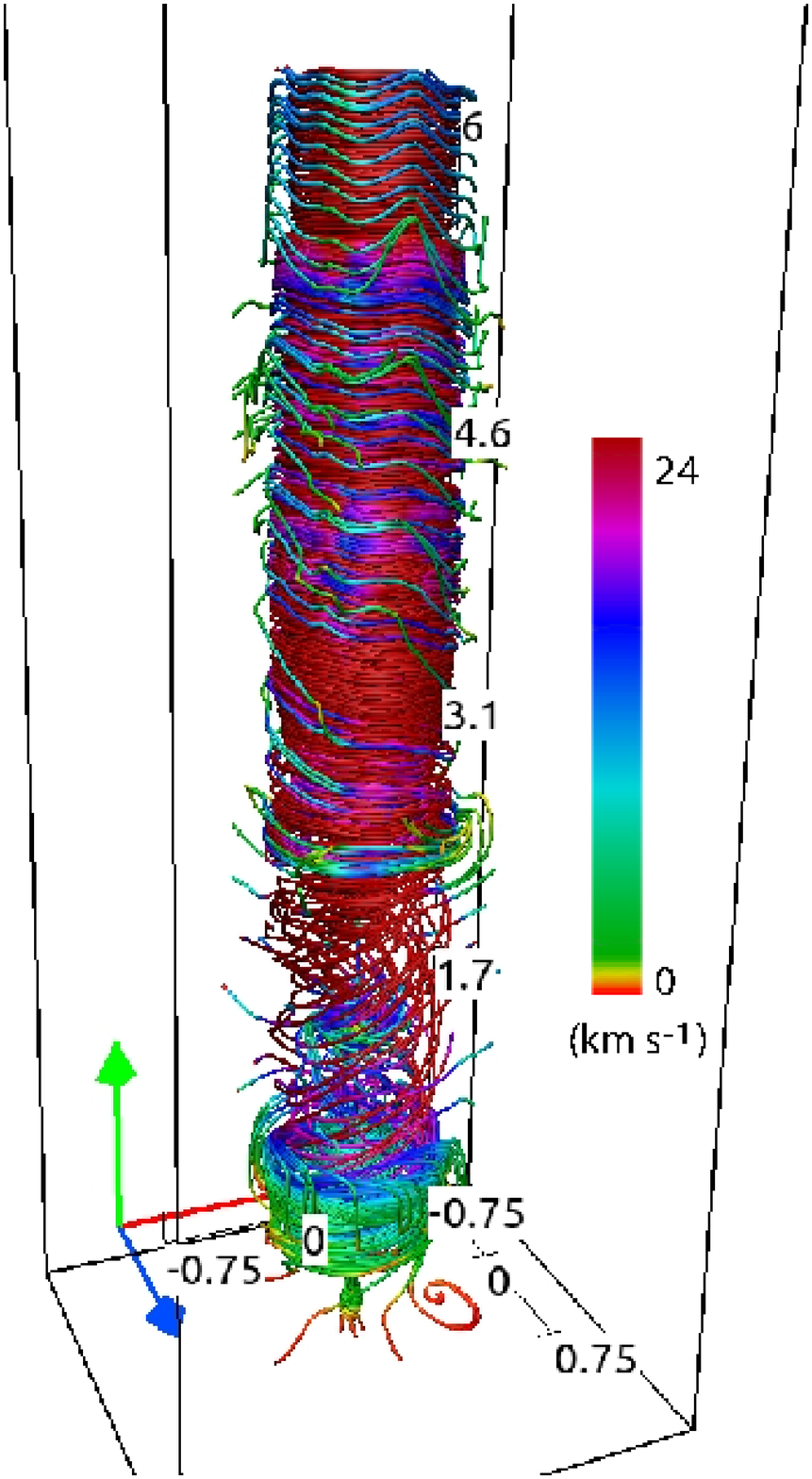}
}
\mbox{
\includegraphics[width=8.5cm,height=9.0cm, angle=0]{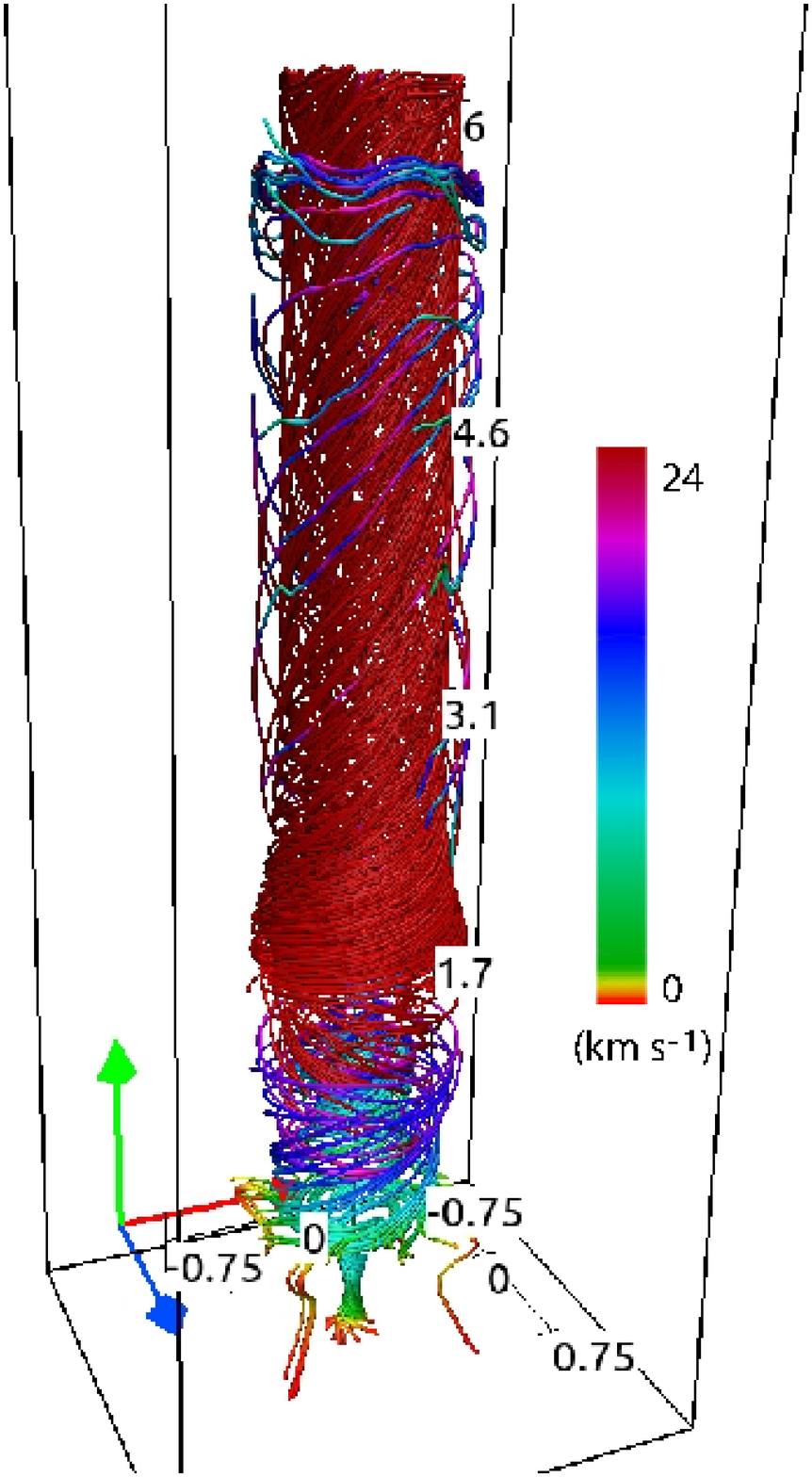}
%\hspace{-1.0cm}
\includegraphics[width=8.5cm,height=9.0cm, angle=0]{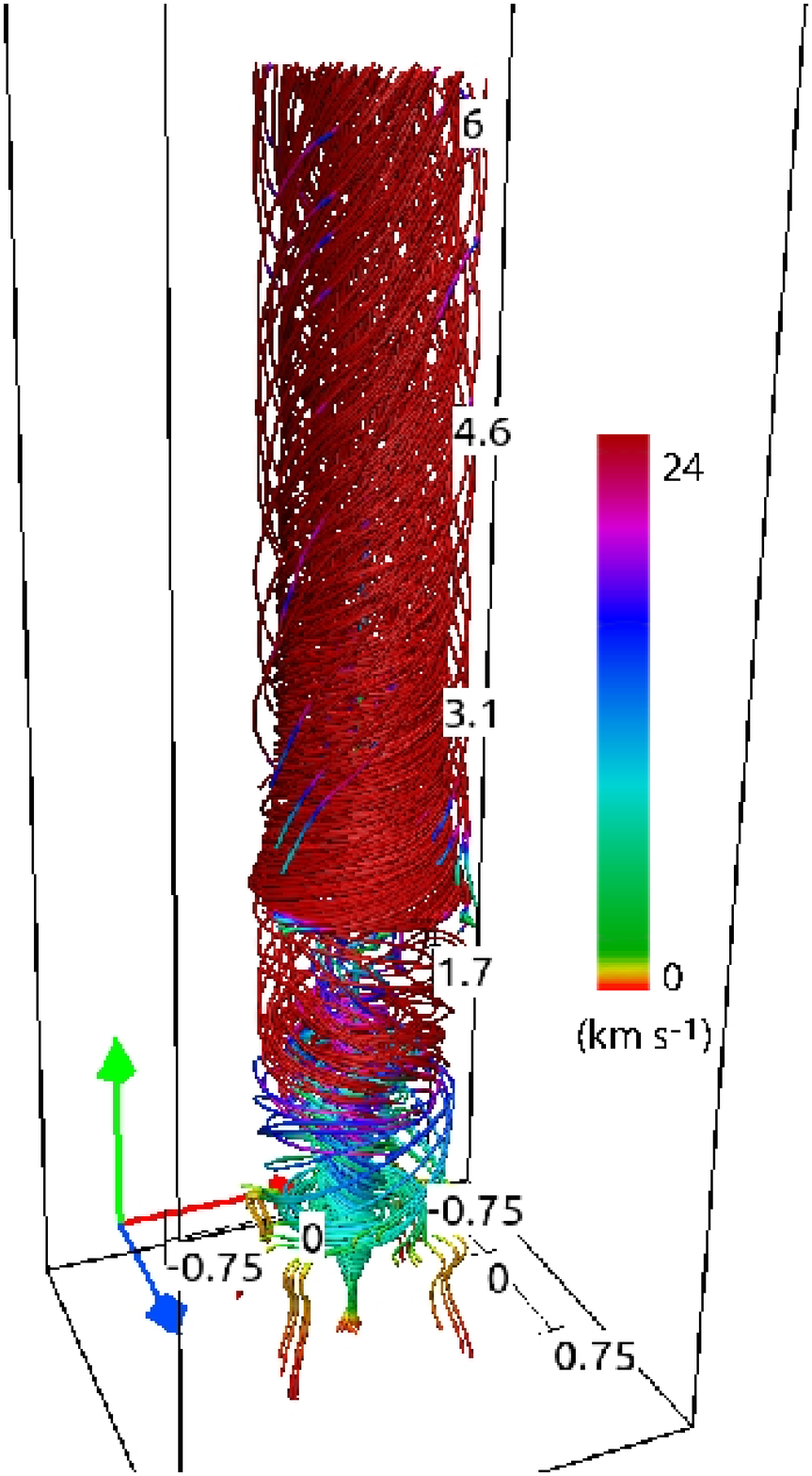}
}
\end{center}
\caption{\small
%Temporal snapshots of 
Streamlines at 
$t=100$ s 
%(left) 
(top-left), $t=150$ s (top-right), 
$t=200$ s (bottom-left), 
and $t=225$ s (bottom-right). 
%$t=250$ s (top) and $t=500$ s (bottom). 
Red, green and blue arrows design the 
directions of the $x$-, $y$-, and $z$-axis, respectively. 
The size of the box shown is 
$(-0.75,0.75)\times (0,6)\times 
(-0.75,0.75)$ Mm. 
% The dark-red lines correspond to the maximum velocity of $\sim 12$ km s$^{-1}$ 
% and the light-blue lines denote the velocity of $\sim 0.1$ km s$^{-1}$. 
The color map corresponds to the 
magnitude of 
a total 
velocity. 
}
\label{fig:V}
\end{figure*}

Figure~\ref{fig:V} shows temporal snapshots of velocity streamlines at $t = 100$ s 
(the left-top panel), $t = 150$ s (the right-top panel), $t = 200$ s (the left-bottom
panel), and $t = 225$ s (the right-bottom panel).  These streamlines are defined by 
the following equations: 
\beq
\frac{dx}{V_{\rm x}} = \frac{dy}{V_{\rm y}} = \frac{dz}{V_{\rm z}} \, .
\eeq 
Note that a signal in the azimuthal velocity is clearly seen in the solar corona. 
Indeed, the streamlines show that velocity perturbations penetrate the transition 
region and enter the solar corona, leading to a generation of the main centrally 
located vortex, which starts developing at $t = 100$~s. 
% This vortex is accompanied by 
% the concentric eddies which are well seen for $t>200$~s (the bottom panels).

Our results clearly 
show that the velocity perturbations are significantly enhanced in the solar 
transition region as a result of very steep temperature gradient. The main
physical consequence of this decrease is that in the inner corona, just above 
the transition region, we see the more velocity streamline perturbations (Figs.~\ref{fig:V}), 
while we do not see much perturbations of the azimuthal component of the 
magnetic fields (cf., Figs. 5-6). 
Note that in the linear limit, $V_{\theta}$ is governed by the wave equation 
of Eq.~(\ref{eq:V_theta}) which was derived by Hollweg (1981). 
% (e.g., 
% Murawski \& Musielak 2010) 
% %
% \beq\label{eq:V_theta}
% \frac{\partial^2 V_{\theta}}{\partial t^2} - c_{\rm A}^2(s) \frac{\partial^2 V_{\theta}}
% {\partial s^2} = 0\, ,
% \eeq
% %
%which 
This equation 
differs from Eq.~(\ref{eq:B_theta}), and this difference affects the evolution 
scenario of velocity perturbations. 

\begin{figure*}
\begin{center}
%\vspace{-2.cm}
\mbox{
\includegraphics[width=8.75cm,height=9.0cm, angle=0]{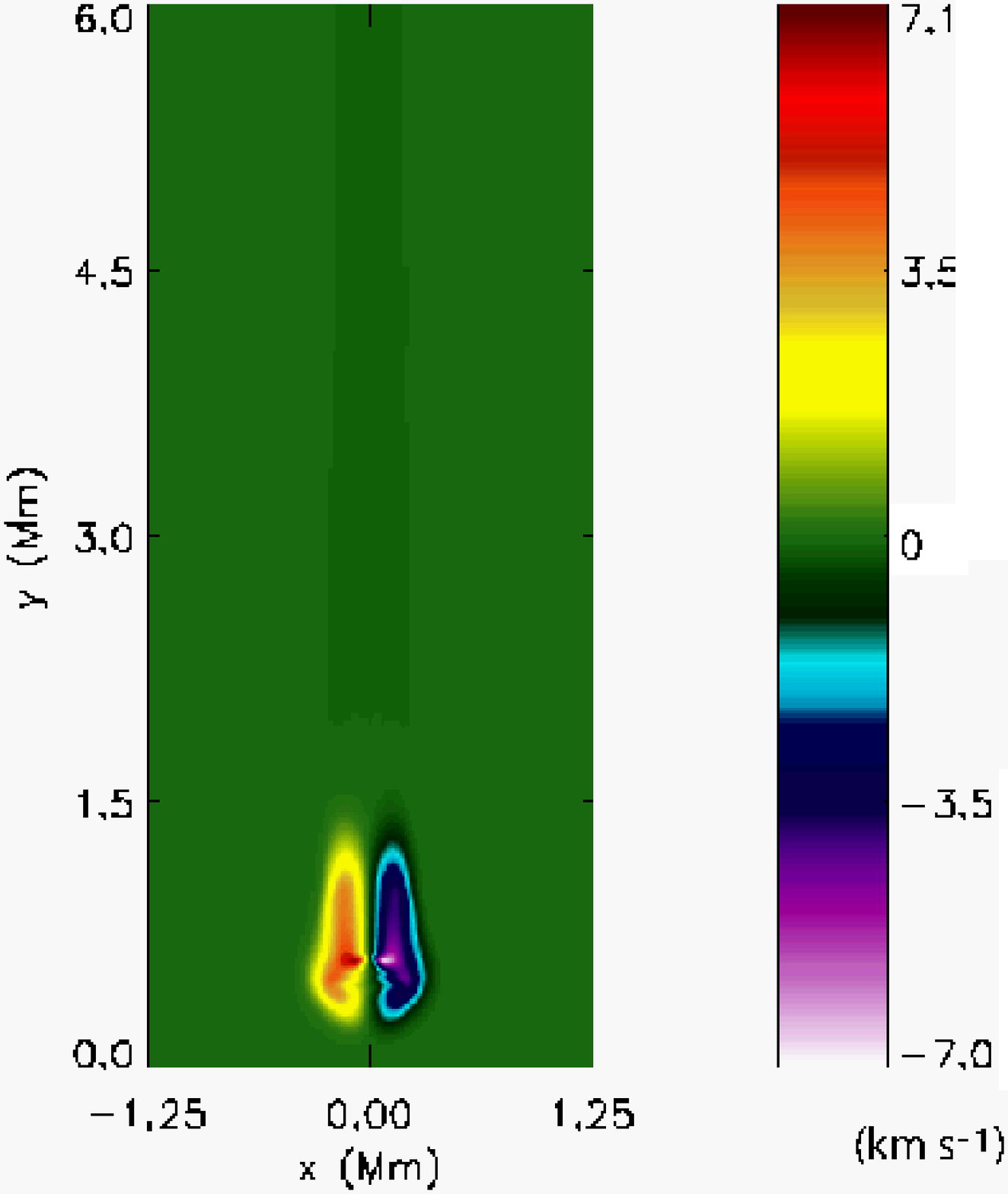}
% \hspace{-0.5cm}
  \includegraphics[width=8.75cm,height=9.0cm, angle=0]{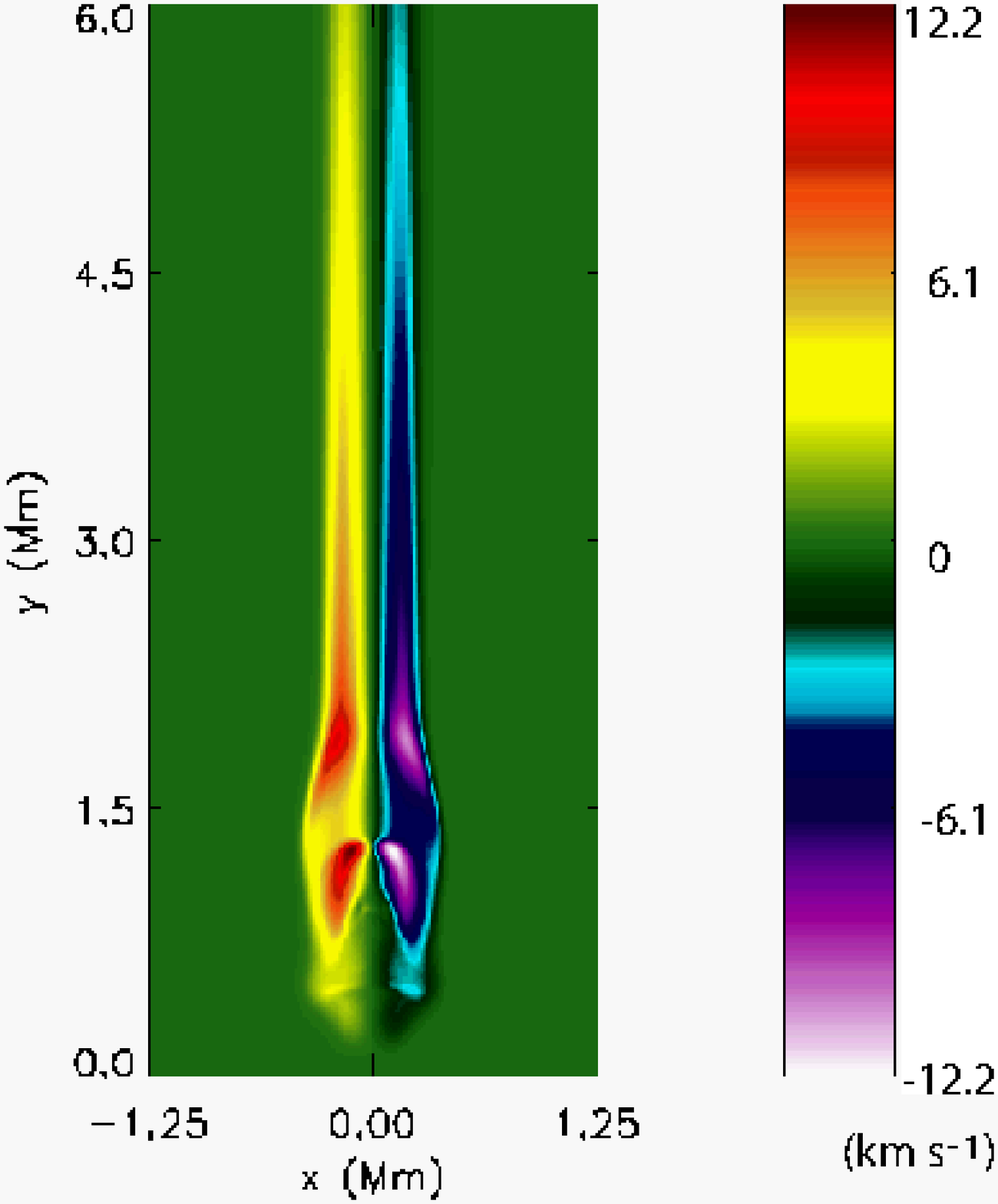}
}
\end{center}
\begin{center}
%\vspace{-1.25cm}
  \mbox{
  \includegraphics[width=8.75cm,height=9.0cm, angle=0]{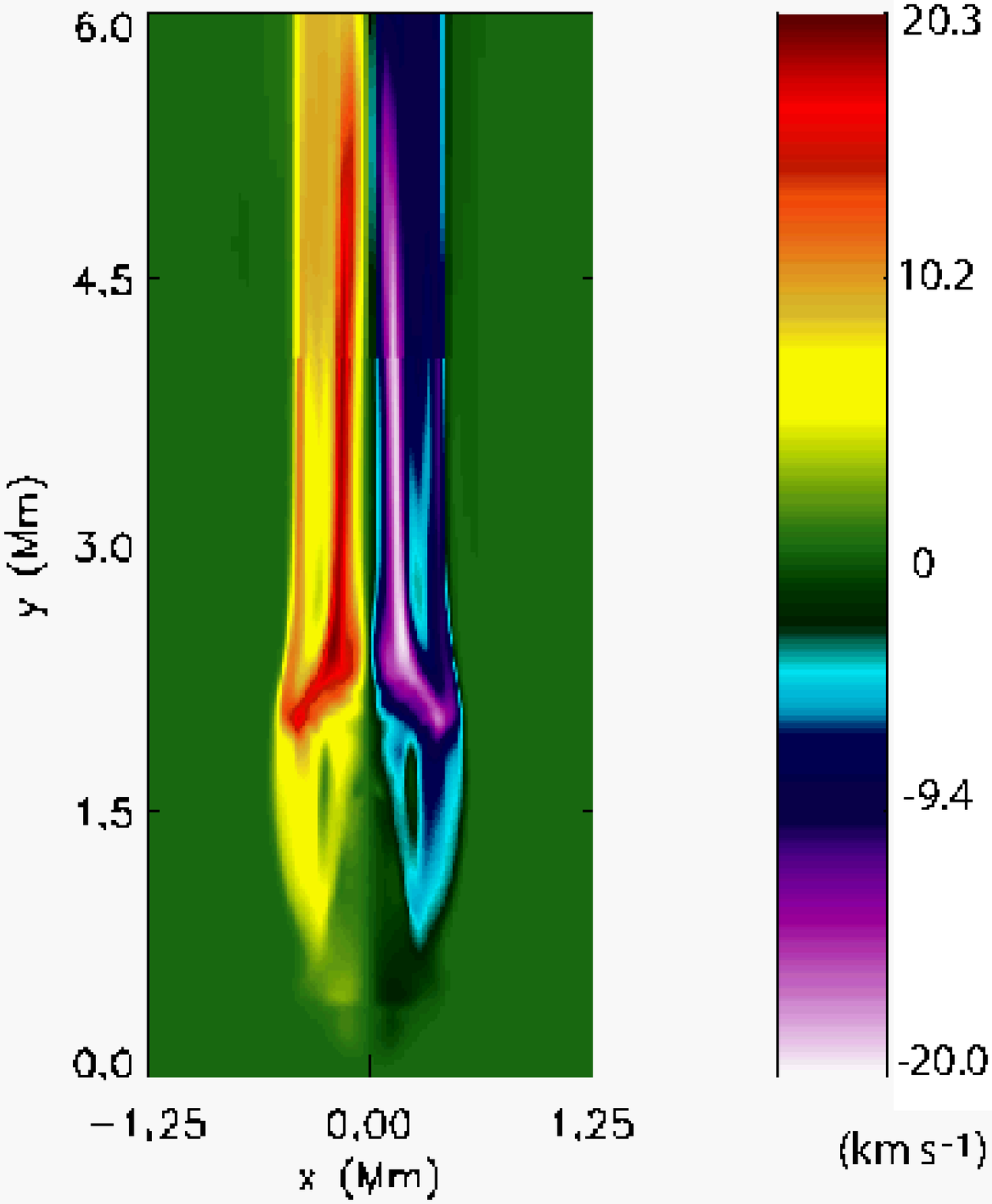}
% \hspace{-0.5cm}
  \includegraphics[width=8.75cm,height=9.0cm, angle=0]{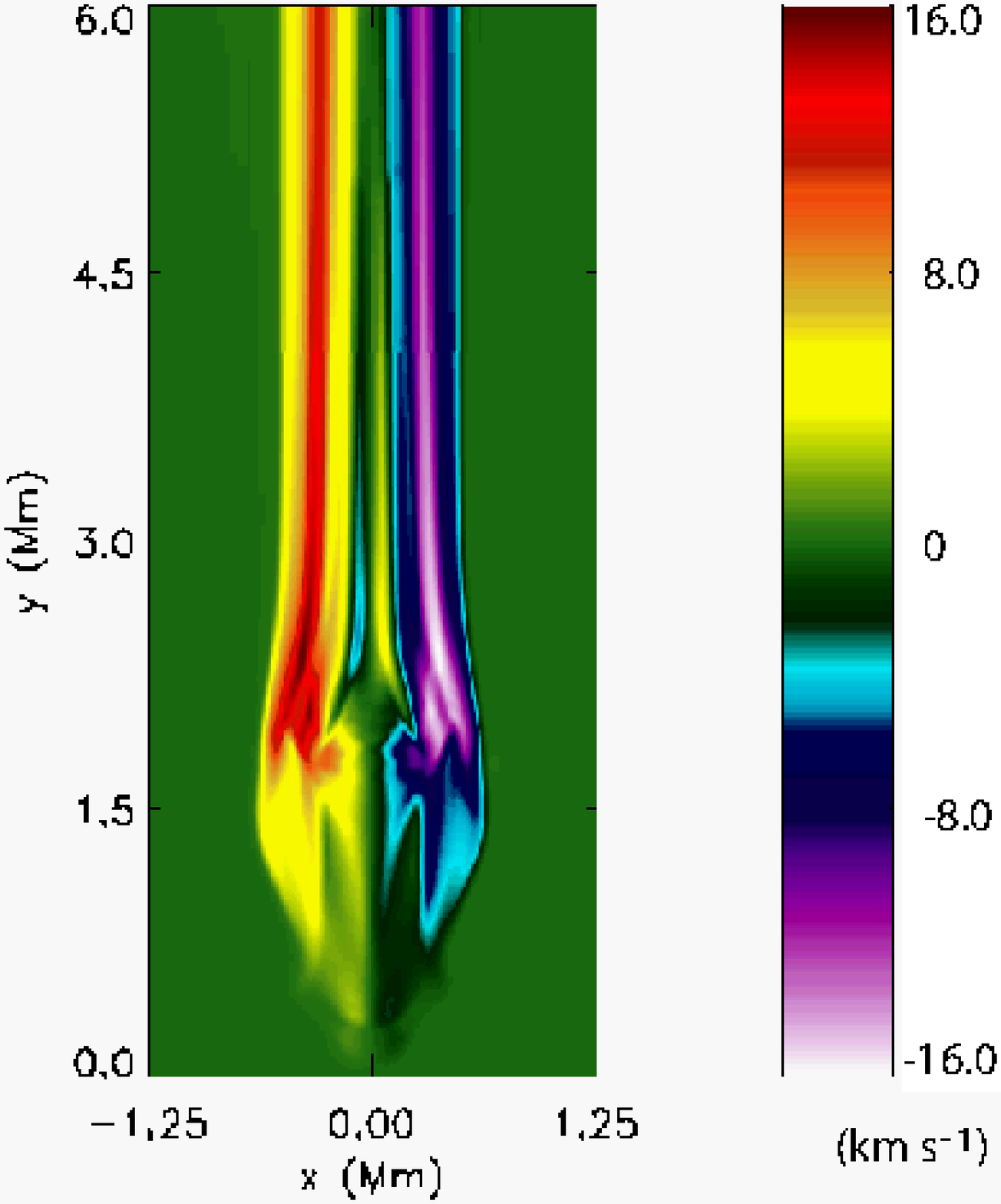}
}
\end{center}
%\vspace{-0.75cm}
\caption{\small
Spatial profiles of $V_{\rm z}(x,y,z=0)$ associated with torsional Alfv\'en waves 
at $t=50$ s (the top-left panel), $t=100$ s (the top-right panel), $t=150$ s (the 
bottom-left panel) and $t=200$ s (the bottom-right panel).  
}
\label{fig:Vz}
\end{figure*}

Torsional Alfv\'en waves can be traced in the 
vertical profiles of $V_{\rm z}(x,y,z=0)=V_{\theta}(x,y,z=0)$ (see Fig.~\ref{fig:Vz}). 
At $t=200$~s (the bottom-right panel), the upwardly propagating Alfv\'en waves resulting 
from the initial pulse are well seen, with their amplitudes of $\vert V_{\rm z}\vert 
\approx 16$ km s$^{-1}$.  The downwardly propagating waves subside as it is 
discernible at $t= 100$~s (the top-right panel) and at $t= 150$~s (the bottom-left panel). 
At $t=100$~s, the upwardly propagating waves reach the 
altitude of more than $y=1.6$ Mm as shown in the top-right panel of the figure. 
%9
% %%%%%%%%%%%%%%%%%%%%%%%%%%%%%%%%%%%%%%%%%%%%%%%%%%%%%%%%%%%%%%%%%%%%%%%%%%%%%%%%%%
%10
% %%%%%%%%%%%%%%%%%%%%%%%%%%%%%%%%%%%%%%%%%%%%%%%%%%%%%%%%%%%%%%%%%%%%%%%%%%%%%%%%%%%%%%%

%%%%%%%%%%%%%%%%%%%%%%%%%%%%%%%%%%%%%%%%%%%%%%%%%%%%%%%%%%%%%%%%%%%%%%%%%%%%%%%%%%%%%%%%%%

As a result of nonlinearity, torsional Alfv\'en waves drive a vertical flow through a 
ponderomotive force 
(Hollweg et al. 1982; 
%e.g., 
%Murawski 2002; 
Nakariakov et al. 1997, 1998). 
%From the $y$-component of Eq.~(\ref{eq:MHD_V}), it follows that 
We refer to Eq.~(6) in Hollweg et al. (1982) which illustrates 
%clearly 
all the terms which contribute
to the component of velocity that is parallel to a magnetic field line, $V_{\parallel}$. 
Denoting by $s$ distance measured along this line and by $r(s)$ a distance of any point 
on this line from the 
%symmetry 
$y$-axis 
%of this line 
we restate
% or referred to with the correct description of all the relevant terms.
Eq.~(6) in Hollweg et al. (1982) as 
\beqa
\nonumber
  \frac{\partial }{\partial t} \left(\frac{\varrho V_{\parallel}}{B_{\parallel}}\right) 
+ \frac{\partial }{\partial s} \left(\frac{\varrho V_{\parallel}^2}{B_{\parallel}}\right) = 
- \frac{1}{B_{\parallel}}\frac{\partial p}{\partial s} + \frac{\varrho g_{\parallel}}{B_{\parallel}} + \\
 \frac{1}{B_{\parallel}} \left[ \left(\varrho V_\theta^2 -\frac{1}{\mu}B_\theta^2\right)\frac{\partial\, \ln r}{\partial s} 
- \frac{\partial }{\partial s} \left(\frac{B_{\theta}^2}{2\mu}\right)
\right]\, ,
\label{eq:poder_force}
%- \frac{1}{2\mu} \frac{\partial B_{\theta}^2}{\partial y}\, .
\eeqa
where $g_{\parallel}$ and $B_{\parallel}$ are the components of, respectively, 
the gravitational acceleration and perturbed magnetic field along the magnetic field line, 
and $V_\theta$ and $B_\theta$ are azimuthal components of velocity and magnetic field, respectively. 
Following Hollweg et al. (1982), we state that the terms in the square bracket result from the $s$-component of 
the Lorentz force: the first term denotes the $s$-component of the centrifugal force that corresponds to 
twisting motions; the second term represents the magnetic tension; the third term is the magnetic pressure. 

\begin{figure}
\begin{center}
% \vspace{-1.5cm}
% \includegraphics[width=9.25cm,height=9.250cm, angle=0]{figs/Vy_t=50.eps}\\
 \includegraphics[width=9.25cm,height=9.250cm, angle=0]{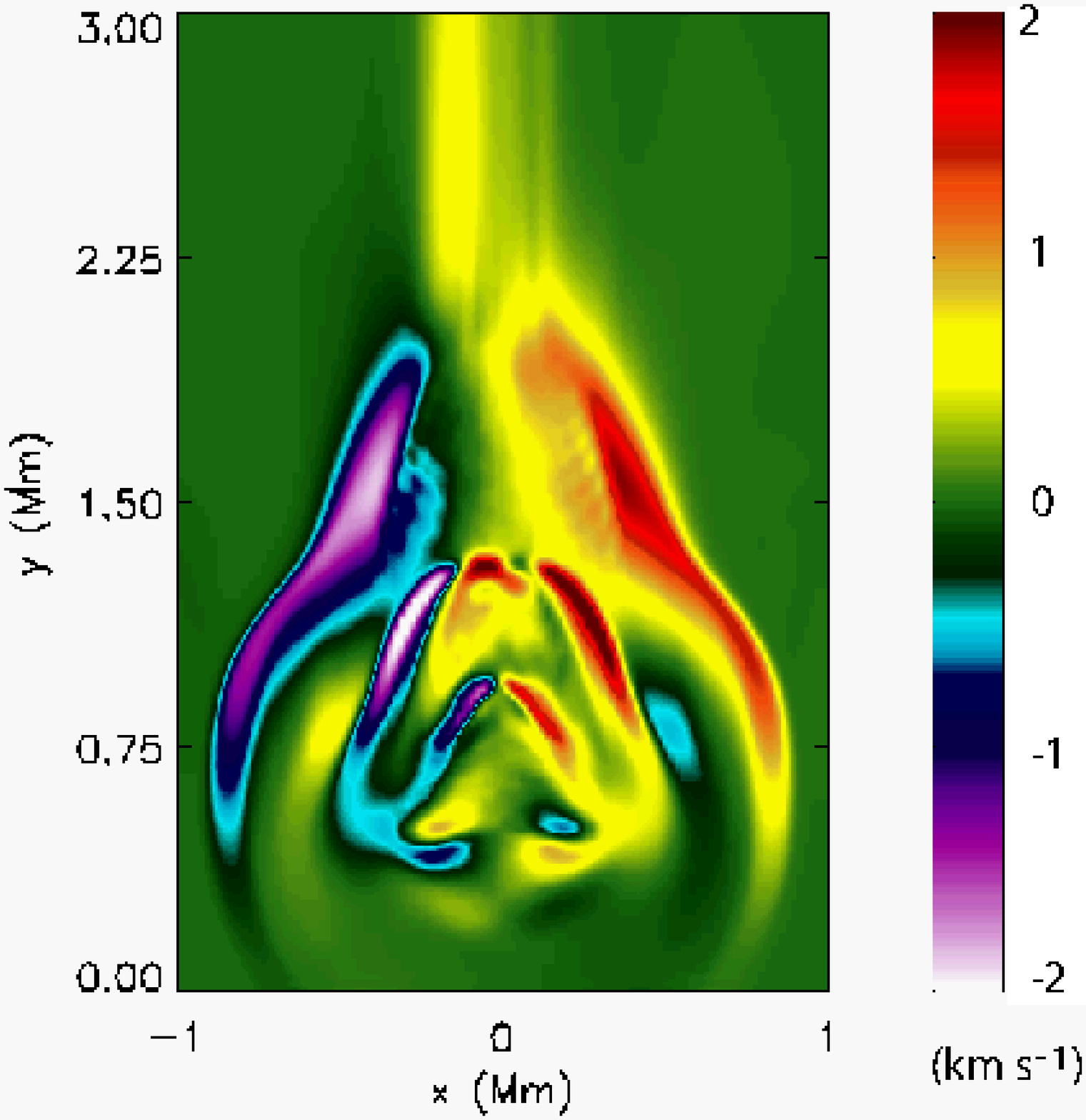}\\
 \vspace{-1.cm}
 \includegraphics[width=9.25cm,height=9.250cm, angle=0]{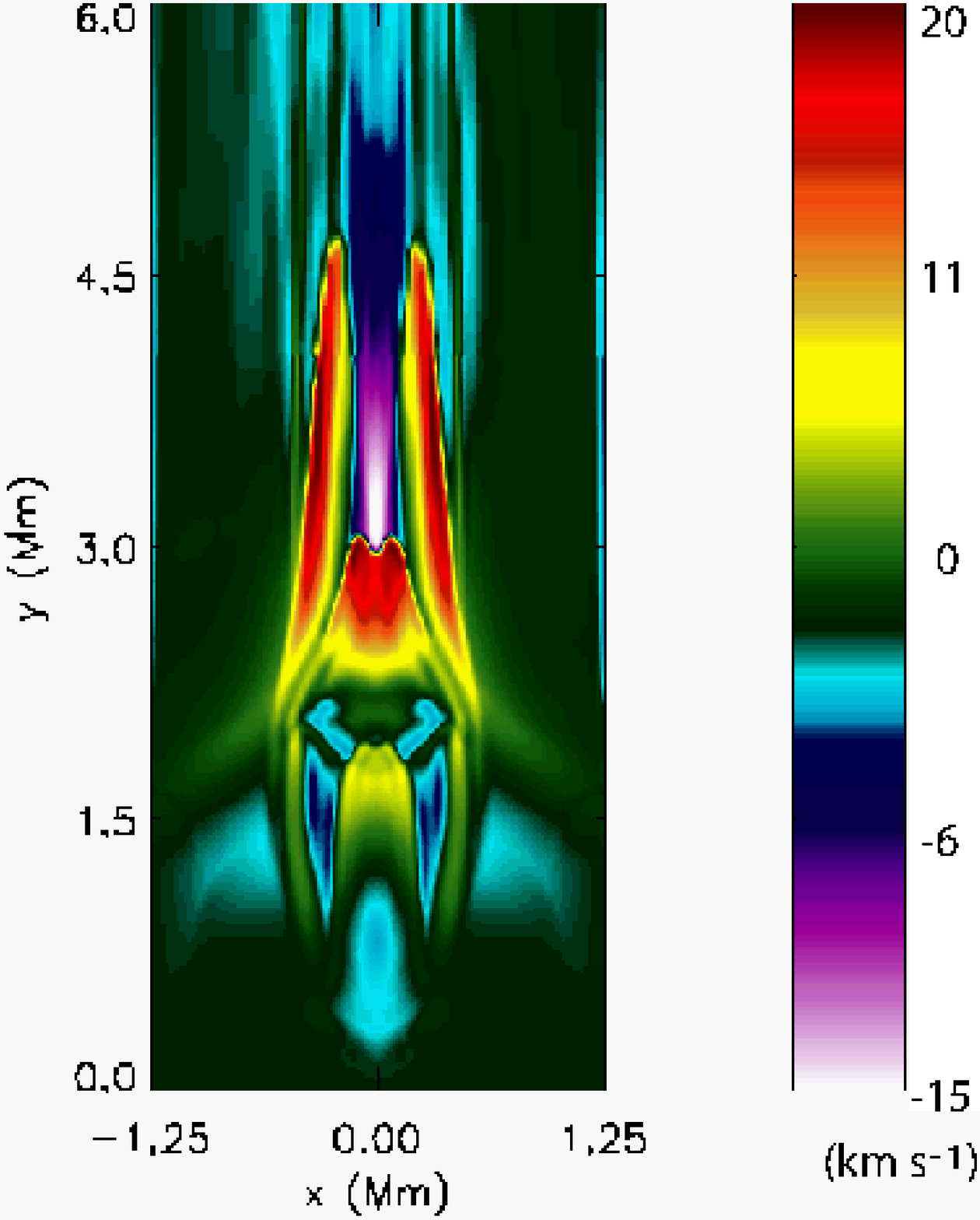}\\
 \vspace{-1.cm}
 \includegraphics[width=9.25cm,height=9.250cm, angle=0]{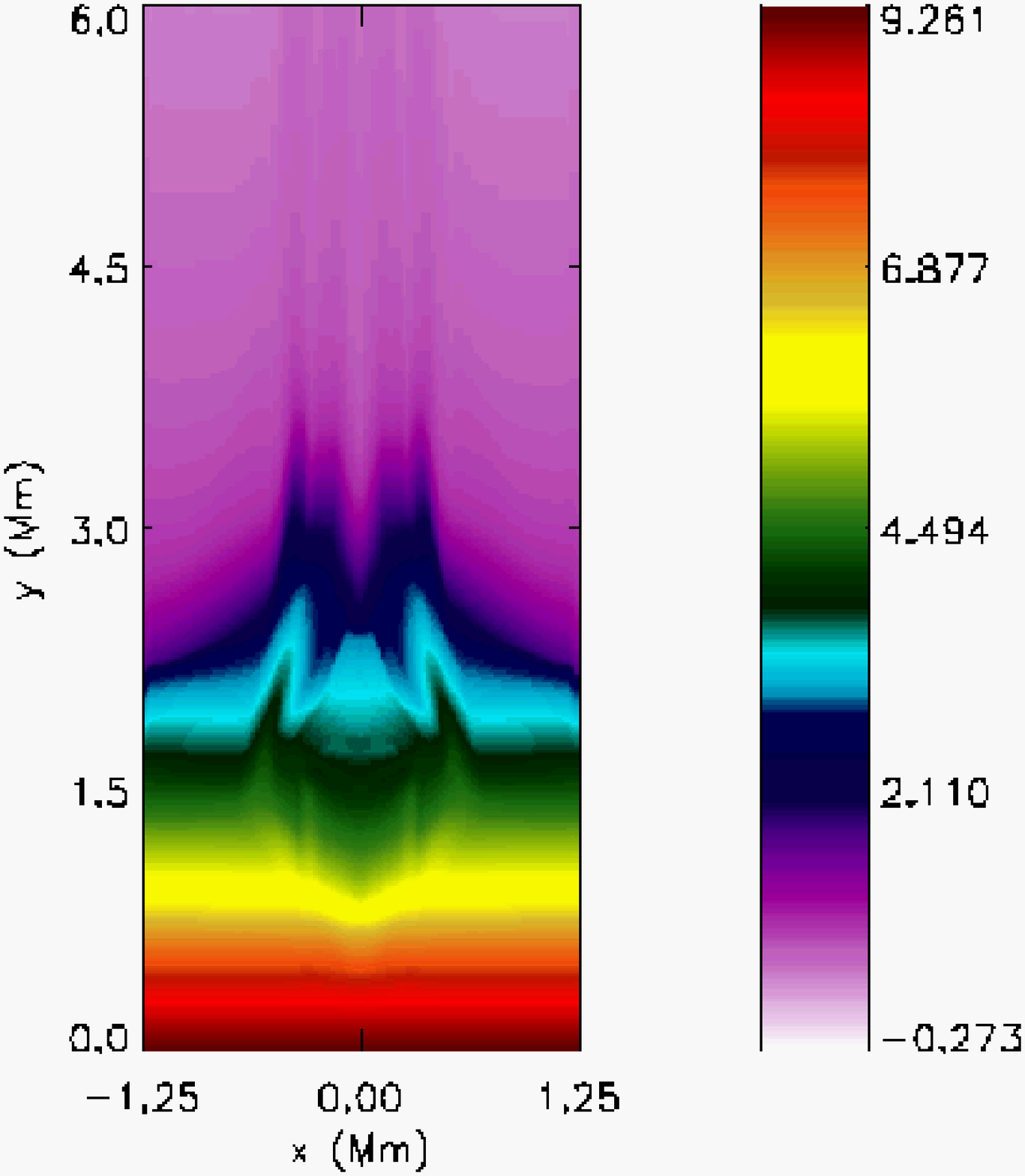}
 \vspace{-1.5cm}
\end{center}
\caption{\small
Spatial profiles of 
$V_{\rm x}(x,y,z=0)$ at 
%$t=50$ s (left-top), 
$t=100$ s (top), 
$V_{\rm y}(x,y,z=0)$ at 
%$t=50$ s (left-top), 
$t=175$ s (middle)
and $\log(\varrho(x,y,z=0)/\varrho_{\rm 0})$ at $t=225$ s (bottom), 
where $\varrho_{\rm 0}=10^{-12}$ kg m$^{-3}$ is the hydrostatic mass density at the level $y=10$ Mm. 
}
\label{fig:Vy}
\end{figure}

The vertical flow can be seen in the spatial profiles of $V_{\rm y}(x,y,z=0)$ (Fig.~\ref{fig:Vy}, middle). 
At $t=400$ s, slow magnetoacoustic waves in a low plasma $\beta$ region of 
the magnetic flux tube of axial symmetry 
as displayed in the right-panel of Fig.~\ref{fig:cs-cA} are weakly coupled to fast 
magnetoacoustic waves, which are described by essentially only $V_{\rm y}$ component of the 
velocity. Our simulations demonstrate that at $t=175$~s slow magnetoacoustic waves already penetrated 
the solar corona, reaching the level higher than $y=4.5$ Mm.  

It is noteworthy 
%to point out 
that non-linear torsional Alfv\'en waves drive perturbations in a mass 
density, which are associated with fast (Fig.~~\ref{fig:Vy}, top) and predominantly slow (Fig.~~\ref{fig:Vy}, middle) magnetoacoustic waves. 
The vertical spatial profile 
of $\log(\varrho(x,y,z=0)/\varrho_{\rm 0})$ is shown at $t=225$ s in the bottom panel of Fig.~\ref{fig:Vy}. 
Here $\varrho_{\rm 0}=10^{-12}$ kg m$^{-3}$ is the hydrostatic mass density at the level $y=10$ Mm. 
At this time the dense plasma jets reach the altitude of $y \approx 3.5$ Mm. The plasma is 
%being 
%then 
lifted up due to the magnetic pressure gradient originating from the Alfv\'en 
waves and acts against the gravity to push the chromospheric material up to the inner solar 
corona. %Such plasma perturbations associated with torsional Alfv\'en waves were recently 
%modeled numerically by Murawski et al. (2014) in the context of a 
%{\bf 
%fast 
%magnetic 
%swirl. 
%twister.
%}
%10
Therefore, such predominant vertical pressure can create the cool plasma jet flows 
typically observed in the solar corona. 
%11
%%%%%%%%%%%%%%%%%%%%%%%%%%%%%%%%%%%%%%%%%%%%%%%%%%%%%%%%%%%%%%%%%%%%%%%%%%%%%%%%%%%%%%%%%%
%%%%%%%%%%%%%%%%%%%%%%%%%%%%%%%%%%%%%%%%%%%%%%%%%%%%%%%%%%%%%%%%%%%%%%%%%%%%%%%%%%%%%%%%%

%12
%%%%%%%%%%%%%%%%%%%%%%%%%%%%%%%%%%%%%%%%%%%%%%%%%%%%%%%%%%%%%%%%%%%%%%%%%%%%%%%%%%%%%%%%%%
\begin{figure}
\begin{center}
 \includegraphics[width=7.0cm,height=6.0cm, angle=0]{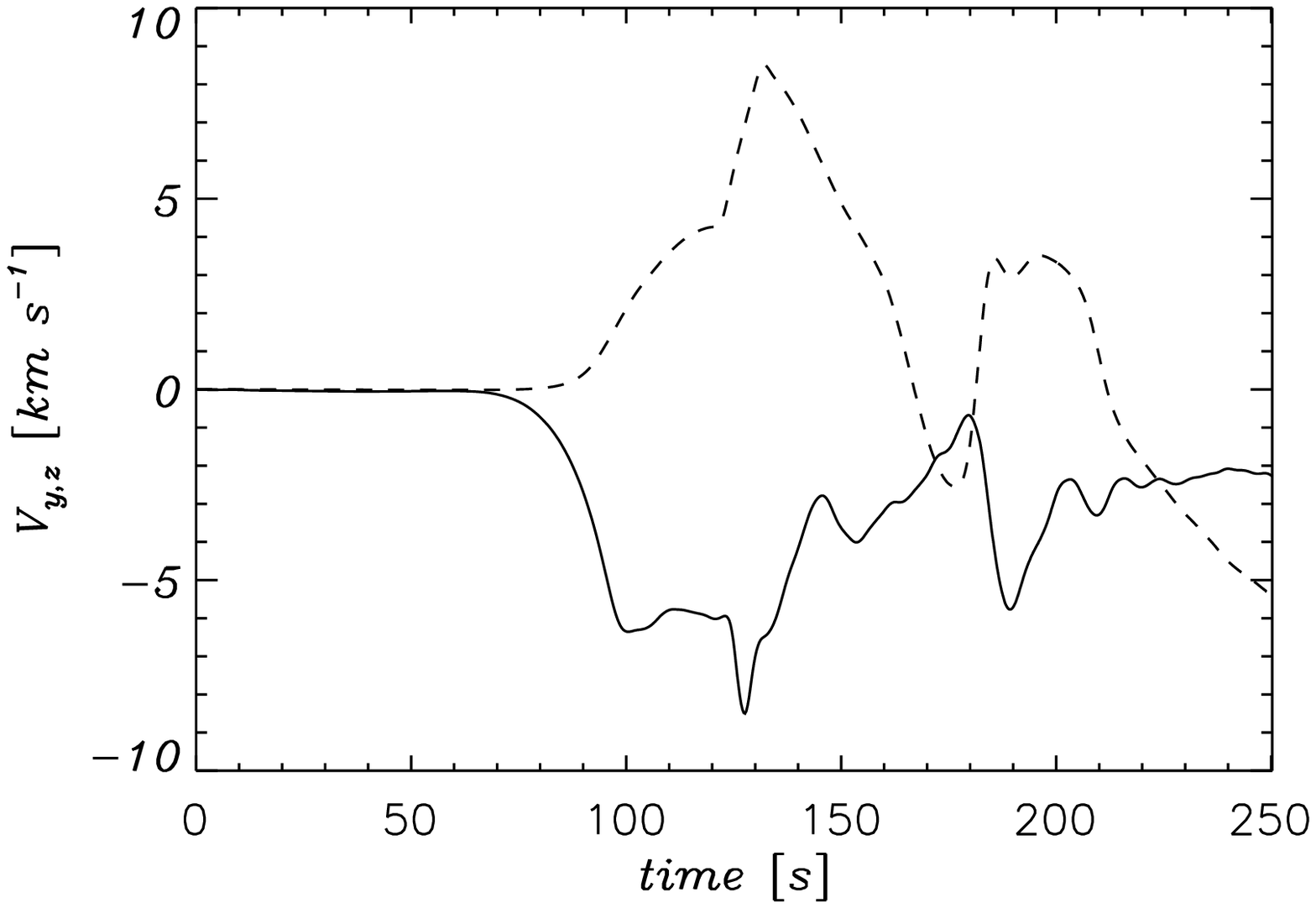}
\end{center}
\caption{\small
Time-signatures of velocities 
%(top): 
$V_{\rm z}$ (solid line) and $V_{\rm y}$ (dashed line),  
%and mass density (bottom), 
collected just below the transition region at the detection point $(x=0.1,y=1.9,z=0.1)$ Mm. 
}
\label{fig:ts-VyVz}
\end{figure}
%%%%%%%%%%%%%%%%%%%%%%%%%%%%%%%%%%%%%%%%%%%%%%%%%%%%%%%%%%%%%%%%%%%%%%%%%%%%%%%%%%%%%%%%%
Figure~\ref{fig:ts-VyVz} displays the time-signatures of $V_{\rm z}$ (the 
solid line) and $V_{\rm y}$ (the dashed line) collected at the detection 
point, $(x=0.2,y=3.5,z=0.2)$ Mm. We infer that slow magnetoacoustic waves, represented by 
$V_{\rm y}$, reach the detection point at $t\approx 130$ s, while Alfv\'en waves arrive 
at this point about $30$ s earlier. 
 
Noting that up to $y\approx 0.7$ Mm Alfv\'en speed is lower than the sound speed 
(Fig.~\ref{fig:cs-cA}, left), the travel times of the torsional Alfv\'en, $t_{\rm At}(y)$, 
and slow magnetoacoustic waves, $t_{\rm st}(y)$, are given by
\beqa
\label{eq:tta}
t_{\rm At}(y) \approx \int_{0.5}^{y} \frac{d {\tilde y}}{c_{\rm A}(r,{\tilde y})}\, ,\\
\label{eq:tts}
t_{\rm st}(y) \approx \int_{0.5}^{y} \frac{d {\tilde y}}{c_{\rm T}(r,{\tilde y})}\, ,
\eeqa
where the integration is performed along the corresponding magnetic field line, and 
$c_{\rm T}=c_{\rm s}c_{\rm A}/\sqrt{c_{\rm s}^2+c_{\rm A}^2}$ is the tube speed. 
Using Eqs.~(\ref{eq:tta}) and (\ref{eq:tts}), we find $t_{\rm At}(y=1.9)\approx 157.7$ 
s and $t_{\rm st}(y=1.9)\approx 173.3$ s. These values are larger than the arrival times of Alfv\'en and slow magnetoacoustic waves. 
This dispatch results from nonlinearity.
%, 

The amplitude of the Alfv\'en waves, seen in $V_{\theta}$, penetrating into the upper region of the atmosphere 
drops from about $V_{\theta}=9$~km s${}^{-1}$ at $t=100$~s just under the transition region, 
%(Fig.~\ref{fig:reflection}, panel $d$) 
through $V_{\theta}=8.4$~km s$^{-1}$ just above the transition region. 
%after about $10$~s 
%(Fig.~\ref{fig:reflection}, panel $h$). 
%and finally disappears. 
We can evaluate the 
%reflection 
transmission 
coefficient,
%
% -----------------------------------------------------------------------------------------.
\beq
\label{eq:cr}
C_{\rm t} = \frac{A_{\rm t}}{A_{\rm i}}\, ,
\eeq
% -----------------------------------------------------------------------------------------.
%
where $A_{\rm i}$ ($A_{\rm t}$) is the amplitude of the incident 
%(reflected) 
(transmitted) 
waves.
Substituting $A_{\rm i}=9$~km~s${}^{-1}$
and $A_{\rm t}=8.4$~km~s${}^{-1}$
into the above formula we get $C_{\rm t}\approx 0.95$,
which means that about than $5$\% of the waves amplitude became reflected
in the transition region and about $95$\% was transmitted into upper atmospheric layers. 
These are rough estimates due to difficulties with precise estimation of wave amplitudes 
which highly vary in time at the transition region. 
%
% The reflection coefficient, $C_{\rm r}$, %=1-C_{\rm r}$,
% vs. initial pulse position $x_{\rm 0}$ for the first reflection from the transition region
% is presented in Fig.~\ref{fig:cr}.
% We expect that the amplitude of the wave signal reflected in the transition region
% is smaller for Alfv\'en waves in larger inclined magnetic field lines,
% which corresponds to larger values of $x_{\rm 0}$.

%
%
\section{Discussion and Conclusions}\label{sec:res_discuss}
In this paper, we present an analytical model of a magnetic flux tube of axial symmetry embedded 
in the solar atmosphere whose photosphere, chromosphere and transition region are 
described by the model of Avrett \& Loeser (2008). Our flux tube model can easily be adopted to any 
axisymmetric magnetic structure by specifying a different magnetic flux. The model was 
used to perform numerical simulations of the propagation of torsional Alfv\'en waves 
and their coupling to magnetoacoustic waves by using the FLASH code.  Torsional Alfv\'en 
waves are launched at the top of the solar photosphere by an initial localized pulse 
in the azimuthal component of velocity.  This pulse mimics the convectively excited 
photospheric vortices or any other kind of plasma rotatory motions recently observed 
in the solar atmosphere (see Section 1 for references and discussion).   

Our results show a complex behavior of torsional Alfv\'en waves and the driven magnetoacoustic 
waves, and their responses in the solar chromosphere, transition region, and inner corona.  
The time and spatial evolution of Alfv\'en waves shown by our numerical results is akin to 
the dynamics of fast solar swirling motions observed in the form of various jets at diverse 
spatio-temporal scales.  Therefore, our model presented here has a potential to describe the 
physics of various observed and recurring swirling motions (jets) in the solar atmosphere in 
which the driver excites torsional Alfv\'en waves as well as the associated plasma perturbations 
(e.g., Wedemeyer-B\"ohm et al. 2012; Murawski et al. 2014, and references cited therein).

Since in our model the 
solar magnetic flux tube of axial symmetry is strongly magnetized in the solar photosphere, 
it can be either associated with strong magnetic bright points (MBPs) (Jess et al. 2009) or with 
the concentration of the magnetic field between intergranular lanes (Wedemeyer-B\"ohm et al. 2012).
With the use of the Swedish Solar Telescope Jess et al. (2009) observed the periodic variations 
above such MBPs in the detected (at different altitudes) Doppler widths of the H$_\alpha$ line.  
The authors claimed that their findings reveal the presence of the propagating torsional Alfv\'en 
waves with their wave periods of $2-12$ min up to the top of the solar chromosphere in the expanding 
wave-guide rooted in the strongly magnetized MBP.  Clearly, our developed model can be used to excite 
torsional Alfv\'en waves and comparing the theoretically predicted range of wave frequencies to that 
established observationally. Indeed, this was one of the main goals achieved in our paper and we 
found the theoretical wave periods to be smaller than $100$ s (Fig.~\ref{fig:ts-VyVz}), which are 
close to the observational findings of Jess at al. (2009). 

Our numerical results also show that the excited perturbations in azimuthal component of the magnetic 
field (Fig.~\ref{fig:Bz}) are seen below the solar transition region. However, in the inner corona 
these perturbations are of a significantly lower amplitude. On the other hand, the perturbations 
in azimuthal component of velocity are stronger in the solar corona.  

Finally, let us point out that our analytical model can easily be adopted to derive the equilibrium 
conditions for any axisymmetric 3D magnetic structure.  Our 
devised numerical model of the solar magnetic flux tube of axial symmetry can be applicable 
to the variety of flux tubes with different photospheric 
fields, and different expansions, as well as with different strengths of the drivers. The model 
can also be used to investigate the nonlinear evolution of coupled torsional Alfv\'en and 
magnetoacoustic waves, and to determine and understand drivers of coronal jets.

{\bf Acknowledgments.} 
We are indebted to the referee for his/her valuable comments and suggestions that 
allowed us to significantly improve the original version of our paper. 
K.M. expresses his gratitude to Zdzislaw Musielak for hospitality during his visit to the 
University of Texas at Arlington in Spring 2014, when a part of this project was done; the
visit as well as the project described in this paper were supported by NSF under the grant 
AGS 1246074 (Z.E.M. \& K.M.). 
The work has also been supported by a Marie Curie International 
Research Staff Exchange Scheme Fellowship within the 7th European Community Framework Program 
(K.M., A.S., \& J.K.). 
In addition, A.S. thanks the Presidium of Russian Academy of Sciences 
for the support in the frame of Program 22 and Russian Foundation of Fundamental Researches 
under the Grant 13-02-00714. 
The software used in this work was in part developed by the 
DOE-supported ASCI/Alliance Center for Astrophysical Thermonuclear Flashes at the University 
of Chicago. The 2D and 3D visualizations of the simulation variables have been carried out 
using respectively the IDL (Interactive Data Language), Python, and VAPOR (Visualization and 
Analysis Platform) software packages.

\end{document}